\documentclass[preprintnumbers,superscriptaddress,nofootinbib,aps,prd,floatfix,twocolumn]{revtex4-2}
\pdfoutput=1
\usepackage{enumerate}
\usepackage{amsmath,amssymb}
\usepackage{graphicx}
\usepackage{slashed}
\usepackage{xspace,slashed}
\usepackage{float}
\usepackage[dvipsnames]{xcolor}
\usepackage{hyperref}
\hypersetup{colorlinks=true, citecolor=Blue, urlcolor=Blue, linkcolor=Blue}
\usepackage[normalem]{ulem}
\usepackage{subfigure,orcidlink}
\usepackage{multirow,array}
\usepackage{braket}
\usepackage{marginnote}
\usepackage{eurosym}
\usepackage{tabularray}
\UseTblrLibrary{booktabs}
\usepackage{bbding}
\usepackage{mathrsfs}

\begin{document}

\title{Forbidden dark matter assisted by first-order phase transition\\and associated gravitational waves}

\author{Satyabrata Mahapatra{\orcidlink{0000-0002-4000-5071}}}
\email{satyabrata@iitgoa.ac.in}
\affiliation{School of Physical Sciences, Indian Institute of Technology Goa, Ponda-403401, Goa, India}

\author{Partha Kumar Paul{\orcidlink{0000-0002-9107-5635}}}
\email{ph22resch11012@iith.ac.in}
\affiliation{Department of Physics, Indian Institute of Technology Hyderabad, Kandi, Sangareddy, Telangana-502285, India.}

\author{Narendra Sahu{\orcidlink{0000-0002-9675-0484}}}
\email{nsahu@phy.iith.ac.in}
\affiliation{Department of Physics, Indian Institute of Technology Hyderabad, Kandi, Sangareddy, Telangana-502285, India.}

%%%%%%%%%%%%%%%%%%%%%%%%%%%%%%%%%%%%%%%%%%%%%%%%%%%%%%%%%%%%
\begin{abstract}
We propose a simple yet testable framework for light fermion dark matter (DM) with mass in the MeV--GeV range, charged under a dark $U(1)_D$ gauge symmetry. Before symmetry breaking, DM annihilates excessively into massless dark gauge bosons, resulting in an under-abundant relic and hence creating severe tension with cosmic microwave background (CMB) and indirect detection constraints. This is naturally remedied by a strongly first-order phase transition (FOPT) in the dark sector, triggered by a scalar field $\Phi$, which gives rise to masses for the dark gauge boson ($X_D$) and the physical scalar ($\phi$).  To achieve the correct relic abundance while evading indirect detection constraints, the $s$-wave annihilation channel $\chi\bar{\chi} \rightarrow X_D \phi$ is strictly maintained in the kinematically forbidden regime. Crucially, due to the $U(1)_D$ gauge structure, the usual $\chi\bar{\chi} \rightarrow \phi\phi$ annihilation is strictly absent at tree level and only arises at the one-loop level. Within this framework, we explore two distinct possibilities. When the loop-induced $\chi\bar{\chi} \rightarrow \phi\phi$ channel is also kinematically forbidden, it predominantly determines the relic density. Conversely, when the $\chi\bar{\chi} \rightarrow\phi\phi$ channel is kinematically allowed, a rich interplay emerges between the tree-level forbidden $\chi\bar{\chi} \rightarrow X_D \phi$ process and the loop-level allowed $\chi\bar{\chi} \rightarrow\phi\phi$ process, with the dominant contribution being dictated sensitively by the DM mass and gauge coupling. The late-time annihilations are highly suppressed in both of these channels, either by the energetic threshold of the forbidden channel or the $p$-wave velocity suppression of the loop-induced $\chi\bar{\chi} \rightarrow\phi\phi$ final state process, rendering the scenario entirely safe from CMB and indirect search bounds. Moreover, the same symmetry-breaking phase transition is strongly first-order, producing a stochastic gravitational wave background that could be probed by upcoming space-based interferometers and pulsar timing arrays. We demonstrate that achieving the observed DM abundance tightly correlates the DM mass with the nucleation temperature of the phase transition. Thus, this setup links the DM relic abundance, dark-sector dynamics, and gravitational wave signals, offering complementary paths to discovery in both terrestrial and cosmological observations.
\end{abstract}
\date{\today}
%%%%%%%%%%%%%%%%%%%%%%%%%%%%%%%%%%%%%%%%%%%%%%%%%%%%%%%%%%%%
\maketitle
%%%%%%%%%%%%%%%%%%%%%%%%%%%%%%%%%%%%%%%%%%%%%%%%%%%%%%%%%%%%%%%%%%%%%%%%%%%%%%%%%%%%%%%%%%%%%%%%%%%%%%%%
\section{Introduction}
\noindent
Understanding the identity of the dark matter (DM), which constitutes approximately 85\% of the matter content of the Universe, remains one of the foremost challenges in modern cosmology. Robust gravitational evidence from galaxy rotation curves, gravitational lensing, and cluster dynamics~\cite{Zwicky:1933gu,Rubin:1970zza,Clowe:2006eq}, together with precision measurements of the cosmic microwave background (CMB)~\cite{Planck:2018vyg,ParticleDataGroup:2020ssz}, indicate that roughly a quarter of the total energy density of the Universe is composed of non-luminous and non-baryonic matter, whose microscopic nature remains unknown.

For several decades, Weakly Interacting Massive Particles (WIMPs) were regarded as the leading DM candidate, motivated by their natural appearance in many extensions of the Standard Model (SM) and their ability to reproduce the observed relic abundance via thermal freeze-out. However, as increasingly sensitive direct detection experiments have continued to report null results, the viability of the canonical WIMP paradigm has been placed under growing tension~\cite{Arcadi:2024ukq,Arcadi:2017kky}. In particular, the absence of statistically significant signals across multiple generations of direct detection experiments has led to increasingly stringent exclusion limits on WIMP–nucleon scattering cross sections. This mounting experimental pressure has motivated the exploration of alternative DM candidates, including light DM with masses $m_{\rm DM}\lesssim 10~\mathrm{GeV}$~\cite{Balan:2024cmq,Cheek:2025nul,Krnjaic:2025noj,Dutta:2019fxn,Essig:2017kqs,Bondarenko:2019vrb,Adhikary:2024btd,Borah:2024yow,Borah:2025wcc,Ma:2025bjf,Ma:2025xoj,Banik:2024zwj,Amiri:2022cbv}. Such sub-GeV to GeV candidates are now being actively probed by a new generation of low-threshold experiments (e.g. SuperCDMS, SENSEI, CRESST, DarkSide, DAMIC), which search for nuclear or electron recoils induced by light DM scattering~\cite{XENON:2019gfn,CRESST:2019jnq,XENON:2024znc,PandaX-II:2021nsg,PandaX:2022xqx,SENSEI:2020dpa,SuperCDMS:2024yiv,DAMIC-M:2025luv}.

However, even light DM produced via conventional thermal freeze-out is subject to severe cosmological and astrophysical constraints. For DM masses below the electroweak scale, thermal freeze-out generically implies sizable annihilation rates that persist well after chemical decoupling. Such late-time annihilations inject energy into the primordial plasma and are therefore tightly constrained by observations of the CMB as well as by indirect detection experiments~\cite{Slatyer:2015jla}. In particular, CMB measurements strongly limit energy injection during the epoch of recombination, excluding standard $s$-wave annihilating DM with masses $m_{\rm DM}\lesssim \mathcal{O}(10)\mathrm{GeV}$ for visible final states~\cite{Slatyer:2015jla,Planck:2018vyg}. Complementary constraints arise from indirect searches for annihilation products in gamma rays and cosmic rays, such as those performed by Fermi-LAT and AMS-02, which rule out the canonical thermal annihilation cross section for DM masses below $\sim 20~\mathrm{GeV}$~\cite{McDaniel:2023bju,Cuoco:2016eej}. Taken together, these results effectively exclude the simplest realizations of light thermal DM.

A viable alternative is the forbidden DM scenarios \cite{DAgnolo:2015ujb}, wherein the dominant annihilation channel is kinematically suppressed at late times.  In a forbidden model, DM annihilates into slightly heavier states during freeze-out, so that annihilation shuts off exponentially as the temperature drops, evading CMB bounds~\cite{DAgnolo:2015ujb,DAgnolo:2020mpt}.  For instance, one can arrange DM to annihilate into heavier dark-sector bosons or scalars, rather than directly into SM particles.  However, if the final states are SM fields, the DM mass and couplings remain highly constrained by laboratory and astrophysical limits.  
We therefore consider a secluded dark sector in which the DM $\chi$ annihilates primarily into a dark gauge boson $X_D$ and a dark Higgs scalar $\phi$ (which breaks the dark $U(1)$ symmetry).  This hidden-sector forbidden annihilation can yield the correct relic density while allowing sizable DM–SM couplings (via gauge kinetic mixing or other portals) that keep the model phenomenologically accessible.  
A key feature of this framework is its reliance on dark gauge symmetry breaking, which shapes the DM phenomenology. Before the phase transition, both the dark gauge boson $X_D$ and the scalar $\phi$ are effectively massless. In this symmetric phase, the DM annihilates excessively into these massless states. If these couplings are large enough to be phenomenologically interesting, this process would completely deplete the relic abundance while remaining highly active at late times, creating severe tension with CMB and indirect detection constraints. A strongly first-order phase transition can resolve this issue. By dynamically generating masses for both $X_D$ and $\phi$, the FOPT introduces a kinematic threshold that alters the available annihilation phase space. Moreover, the sizable couplings required to drive this strong FOPT naturally generate a stochastic gravitational wave (GW) background that lies within the sensitivity of present and future GW detectors \cite{Hindmarsh:2020hop,Athron:2023xlk,Mahapatra:2025vzu,Ma:2026tyk,Feng:2024pab,Feng:2025wvc}. 

Following symmetry breaking, the relic abundance is governed by an interesting interplay of annihilation channels. Because the strong FOPT dynamics naturally dictate a hierarchical mass spectrum ($M_{X_D} \gg M_\phi$), the $\chi\bar{\chi} \rightarrow X_D X_D$ channel is severely phase-space suppressed. Consequently, the tree-level, $s$-wave process $\chi\bar{\chi} \rightarrow X_D \phi$ naturally dominates the thermal cross-section. To achieve the correct relic density while avoiding late-time constraints, we deliberately tune the mass splitting such that $2M_\chi < M_{X_D} + M_\phi$. This restricts the dominant $s$-wave channel to the thermally-assisted ``forbidden" regime during freeze-out. Furthermore, due to the specific $U(1)_D$ gauge charge assignments in our model, the $\chi\bar{\chi} \rightarrow \phi\phi$ annihilation is strictly absent at tree level. It manifests only as a one-loop process (e.g., via a triangle diagram and box diagram with gauge boson and DM running in the loop) that is intrinsically $p$-wave suppressed.

With the dominant tree-level $\chi\bar{\chi} \rightarrow X_D \phi$ process strictly residing in the forbidden regime, the relic density determination bifurcates into two distinct phenomenological cases. In the first case, the loop-induced $\chi\bar{\chi} \rightarrow\phi\phi$ channel is also kinematically forbidden, leaving the relic abundance exclusively determined by the forbidden $\chi\bar{\chi} \rightarrow\phi \phi$ process even if it suffers from loop effect and $p-$wave suppression. In the second case, the $\chi\bar{\chi} \rightarrow\phi\phi$ channel is kinematically allowed, leading to a rich interplay between the heavily Boltzmann-suppressed tree-level $\chi\bar{\chi} \rightarrow X_D \phi$ process and the phase-space open (but loop- and $p$-wave suppressed) $\chi\bar{\chi} \rightarrow\phi\phi$ process. The channel that ultimately dictates the relic density depends sensitively on the DM mass and the gauge coupling. Crucially, in all cases, late-time annihilations are highly suppressed, either by the energetic thresholds of the forbidden mechanism or by the velocity dependence of the $p$-wave loop process, thus keeping the scenario completely safe from CMB and indirect search.

In summary, this scenario not only achieves the correct thermal relic density but also offers multiple experimental avenues, including direct detection of DM, searches for new forces, as well as gravitational wave astronomy. Moreover, the model predicts a striking correlation between the DM mass and the gravitational wave spectrum: heavier DM requires a higher-scale phase transition to maintain the forbidden mechanism, shifting the GW peak frequency to higher values. This robust model prediction provides a powerful consistency check, linking the microscopic properties of DM to distinct signatures in gravitational wave detectors.

The rest of the paper is built up as follows. In section \ref{sec:fdm}, we discuss the forbidden DM. The minimal model is introduced in section \ref{sec:model}, followed by the First-order phase transition, which is discussed in section \ref{sec:fopt}. We discuss the forbidden DM from the first-order phase transition in section \ref{sec:fdmfopt}. {We discuss other phenomenological constraints in section \ref{sec:constraint}.} Section \ref{sec:gwfopt} discusses gravitational wave production from the first-order phase transition. We finally conclude in section \ref{sec:concl}.

%%%%%%%%%%%%%%%%%%%%%%%%%%%%%%%%%%%%%%%%%%%%%%%%%%%%%%%%%%%%%%%%%%%%%%%%%%%%
\section{Forbidden dark matter}\label{sec:fdm}
\noindent
We consider a scenario where a Dirac fermionic DM $\chi$ couples to dark sector states $\mathscr{D}_1$ and $\mathscr{D}_2$, through an effective interaction term $\mathcal{L_{\rm int}} \supset g_{\rm eff} ~\bar{\chi}~\chi~ \mathscr{D}_1 \mathscr{D}_2 $. The relic abundance is determined through the annihilation process $\chi \bar{\chi} \rightarrow \mathscr{D}_1 \mathscr{D}_2$, with a thermally averaged cross-section of the order $\langle \sigma v \rangle \sim g_{\rm eff}^2 / M_\chi^2$, where we assume that $\mathscr{D}_1$ and $\mathscr{D}_2$ are massless or have negligible mass compared to $\chi$. This can reproduce the observed DM abundance for weak-scale masses and a sizable coupling. For lighter DM below the weak scale, however, the annihilation cross-section becomes too large, driving the relic density to under-abundant values unless $g_{\rm eff} \ll 1$. An attractive alternative is to consider annihilation into slightly heavier final states, the so-called forbidden channels. In this case, the annihilation is kinematically forbidden at zero temperature but proceeds in the early Universe thanks to the Boltzmann tail, leading to an exponentially suppressed cross-section at late times. This allows light DM to achieve the correct relic density even with order-one couplings~\cite{DAgnolo:2015ujb,DAgnolo:2020mpt}.

In the forbidden regime, the thermally averaged annihilation cross-section can be related to the reverse process by detailed balance,
\begin{eqnarray}
\langle\sigma_{\chi \bar{\chi} \rightarrow \mathscr{D}_1 \mathscr{D}_2} v\rangle=\frac{n^{\rm eq}_{\mathscr{D}_1}n^{\rm eq}_{\mathscr{D}_2}}{(n^{\rm eq}_{\chi})^2} \langle\sigma_{ \mathscr{D}_1 \mathscr{D}_2\rightarrow\chi \bar{\chi}} v\rangle,
\end{eqnarray}
where $n^{\rm eq}_i$
denotes the equilibrium number density of species $i$, which for non-relativistic particles is given by $n^{\rm eq}_i=g_i(M_iT/2\pi)^{3/2}e^{-M_i/T}$. Defining $r_i = M_{\mathscr{D}_i}/M_\chi\,, \Delta = M_{\mathscr{D}_1}+M_{\mathscr{D}_2}-2M_\chi)/2M_\chi$ and $x=M_\chi/T$, the ratio of equilibrium densities yield, 
$$ \frac{n^{\rm eq}_{\mathscr{D}_1}n^{\rm eq}_{\mathscr{D}_2}}{(n^{\rm eq}_{\chi})^2}\simeq\frac{g_{_{\mathscr{D}_1}}g_{_{\mathscr{D}_2}}}{g_\chi^2}(r_1r_2)^{3/2}e^{-2\Delta x}$$
where the exponential factor $e^{-2 \Delta x}$ encodes the kinematic suppression from the heavier final states. Including the phase-space factor for the reverse process, the thermally averaged forbidden annihilation cross-section can be written schematically as
\begin{eqnarray}
\langle\sigma_{\chi \bar{\chi} \rightarrow \mathscr{D}_{1} \mathscr{D}_2} v\rangle&=& \frac{g_{_{\mathscr{D}_1}}g_{_{\mathscr{D}_2}}}{g_\chi^2}\mathcal{F}(r_1,r_2)e^{-2\Delta x}  \langle\sigma_{ \mathscr{D}_1\mathscr{D}_2\rightarrow\chi \bar{\chi}} v\rangle,
\end{eqnarray}
where $\mathcal{F}(r_1,r_2)$ is a kinematic function, which for nearly degenerate final states $M_{\mathscr{D}_1}\simeq M_{\mathscr{D}_2}$, reduces to the familiar $(1+\Delta)^{3/2}$ factor in the literature. The key point is that the annihilation rate is exponentially suppressed, so that the resulting relic abundance scales approximately as $\Omega_{\chi}h^2\propto e^{2\Delta x_f}$ with $x_f= M_{\chi}/T_{f}$, the freeze-out parameter. As a result, even relatively light DM with sizable couplings can reproduce the observed relic density by appropriately tuning the mass splitting $\Delta$.

%%%%%%%%%%%%%%%%%%%%%%%%%%%%%%%%%%%%%%%%%%%%%%%%%%%%%%%%%%%%%%%%%%%%%%%%%%
\section{The minimal model}\label{sec:model}
\noindent
We now introduce a minimal setup that realizes the forbidden DM scenario triggered by a first-order phase transition in the dark sector. The model contains a Dirac fermion $\chi$ with charge $+1$ under a $U(1)_D$ gauge symmetry and a complex scalar singlet $\Phi = (\phi + i \eta)/\sqrt{2}$ with charge $-1$. The relevant dark sector Lagrangian is given by
\begin{eqnarray}
\mathcal{L}_{U(1)_D}&=&\bar{\chi}(i\gamma^\mu \mathcal{D}_\mu-M_{\chi})\chi-\frac{1}{4}X_{\mu\nu}X^{\mu\nu}-\frac{\epsilon}{2}B_{\mu\nu}X^{\mu\nu}\nonumber\\&&+(\mathcal{D}_\mu\Phi)^*(\mathcal{D}_\mu\Phi)+\mu_{\Phi}^2\Phi^*\Phi-\lambda_\Phi(\Phi^*\Phi)^2,
\end{eqnarray}
where the covariant derivative is defined as $\mathcal{D}_\mu=\partial_\mu+i Q_{D} \textsl{g}_{D}(X_D)_\mu$with $Q_D$ being the charge of the particle under $U(1)_D$ symmetry, $\epsilon$ is the kinetic mixing parameter between the dark gauge field $(X_D)_\mu$ and the hypercharge gauge boson $B_\mu$. After spontaneous breaking of the $U(1)_D$ symmetry, the scalar can be parameterized around its vacuum expectation value (VEV) as
\begin{eqnarray}    \Phi=\frac{\phi+v_\phi}{\sqrt{2}},
\end{eqnarray}
where $v_\phi$ denotes the VEV of the dark scalar field. The symmetry breaking generates masses for the dark gauge boson and the physical scalar $\phi$ as 
\begin{eqnarray}
    M_{X_D}=\textsl{g}_{D} v_\phi,~M_{\phi}=\sqrt{2\lambda_\Phi}v_\phi.
\end{eqnarray}

\begin{figure}[h]
\centering
\includegraphics[width=0.45\textwidth]{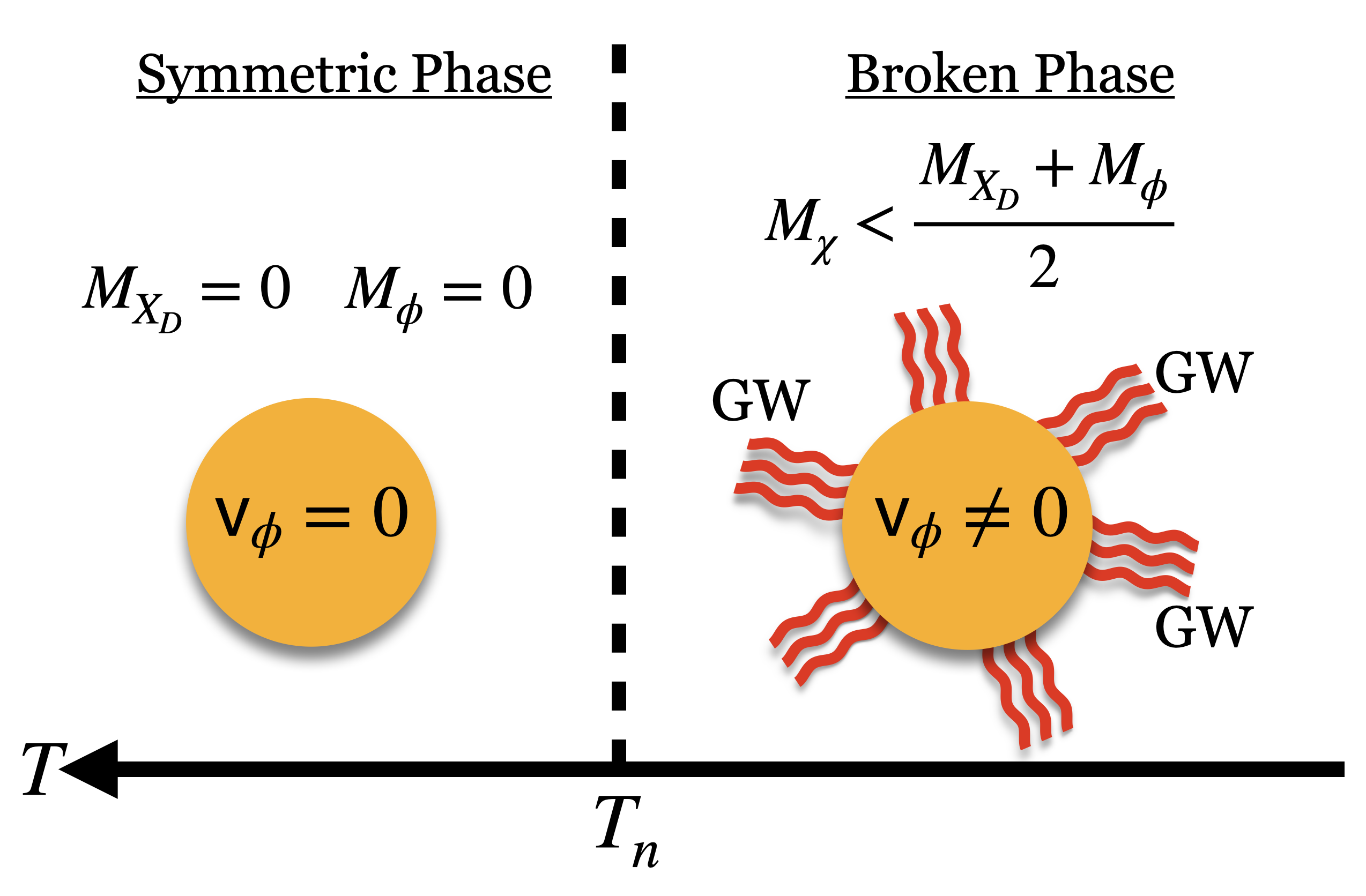}
\caption{Pictorial representation of our model, which gives rise to gravitational waves via a first-order phase transition and correct relic of DM through the forbidden channel. }
\label{fig:pic}
\end{figure}

A pictorial summary of the proposed scenario is shown in Fig.~\ref{fig:pic}, which captures the essential thermal history of the dark sector in this framework. In the high-temperature symmetric phase, the dark scalar field sits at the origin, $v_\phi = 0$, so that both the dark gauge boson and the scalar remain massless and the forbidden annihilation channel is not operative. As the temperature drops below the nucleation temperature $T_n$, the dark sector undergoes a first-order phase transition, and the scalar acquires a non-zero vacuum expectation value, $v_\phi \neq 0$, thereby generating masses for both the dark gauge boson and the scalar itself.

In the broken phase, the mass hierarchy is arranged such that $M_\chi < (M_{X_D} + M_\phi)/2$, rendering the process $\chi \bar{\chi} \to X_D \phi$ a forbidden channel at late times while still allowing it to proceed efficiently in the early Universe due to the thermal population of the heavier final states. This realizes the forbidden DM mechanism discussed above and enables the correct relic abundance for light DM with sizable couplings while simultaneously producing a stochastic gravitational wave background from the strong first-order phase transition. Taken together, this setup tightly links the origin of the DM relic density to testable signals in upcoming gravitational wave experiments and dark sector searches. 
Prior to analyzing the DM phenomenology, we first outline the dynamics of the dark sector phase transition and determine the regions of parameter space that yield a strongly first-order transition. 

%%%%%%%%%%%%%%%%%%%%%%%%%%%%%%%%%%%%%%%%%%%%%%%%%%%%%%%%%%%%%%%%%%%%%%%%%%%%
\section{First-order phase transition}\label{sec:fopt}
\noindent
As emphasized above, the same dark sector dynamics that generate the correct relic abundance through a forbidden channel is facilitated by a strong first-order phase transition, providing a complementary source of a stochastic gravitational wave background. In this section, we focus on the thermal evolution of the scalar field and identify regions of parameter space in which $U(1)_D$ symmetry breaking proceeds via a strongly first-order phase transition, suitable for generating potentially observable gravitational wave signals.

The scalar potential receives thermal corrections at finite temperature, and it can be written as
\begin{eqnarray}
V_{\rm eff}(\phi,T)&=&V_{\rm tree}(\phi)+V_{\rm CW}(\phi)+V_{\rm ct}(\phi)+V_{T}(\phi,T)\nonumber\\&&+V_{\rm daisy}(\phi,T),
\end{eqnarray}
where $V_{\rm tree}(\phi)$ is the tree level potential, $V_{\rm CW}(\phi)$ is the Coleman-Weinberg one-loop correction, $V_{\rm ct}(\phi)$ is the counter term, $V_T(\phi)$ is the finite temperature contribution, and $V_{\rm daisy}$ accounts for the daisy correction. The explicit expressions of these terms are given in Appendix \ref{app:fopt}. At high temperatures, the effective potential is minimized at $\phi=0$. As the Universe cools and the temperature drops, the shape of the potential evolves, and a second, energetically favorable minimum at $\phi\neq 0$ can appear, separated from the origin by a potential barrier. In this situation, the system undergoes a first-order phase transition (FOPT), in which bubbles of the true vacuum nucleate within the metastable false vacuum and subsequently expand. 

Such an FOPT can efficiently produce gravitational waves via three main processes: bubble wall collisions, sound waves in the plasma, and magnetohydrodynamic turbulence in the plasma. The corresponding spectra and their combination into the total gravitational wave signal are discussed in detail in Appendix~\ref{app:gw}. For phenomenological purposes, the gravitational wave signal is largely characterized by two key parameters evaluated at the nucleation temperature $T_n$: the strength of the transition, quantified by the ratio of the released vacuum energy to the radiation energy density, and the inverse duration of the transition in units of the Hubble rate.
These are conventionally expressed as
\begin{eqnarray}
    \alpha(T_n)=\frac{\rho_{\rm vac}(T_n)}{\rho_{\rm rad}(T_n)},
\end{eqnarray}
where
\begin{eqnarray}
\rho_{\rm vac}(T)&=&V_{\rm eff}(\phi_{\rm false},T)-V_{\rm eff}(\phi_{\rm true},T)\nonumber\\&&-
T\frac{\partial}{\partial T}\left[V_{\rm eff}(\phi_{\rm false},T)-V_{\rm eff}(\phi_{\rm true},T)\right],
\end{eqnarray}
and
\begin{eqnarray}
\rho_{\rm rad}(T)=\frac{\pi^2}{30}g_{*}T^4,
\end{eqnarray}
with $g_*$ denoting the number of relativistic degrees of freedom (d.o.f) at $T=T_n$. The parameter $\alpha(T_n)$ thus measures how strongly the phase transition energy budget departs from that of a purely radiation-dominated Universe. The timescale of the transition is encoded in the quantity $\beta$, defined as the inverse duration of the phase transition and commonly expressed in units of the Hubble rate at nucleation,
\begin{eqnarray}
\frac{\beta}{\mathcal{H}(T_n)}\simeq T_n\frac{d}{dT}\bigg(\frac{S_3}{T}\bigg)\bigg|_{T=T_n}.
\end{eqnarray}
where $S_3$ is the three-dimensional Euclidean action of the critical bubble configuration. A smaller value of $\beta/\mathcal{H}(T_n)$ corresponds to a slower, more prolonged transition and typically enhances the gravitational wave signal, while larger values indicate a faster transition with a more suppressed signal.

\begin{figure}[h]
\centering
\includegraphics[width=0.48\textwidth]{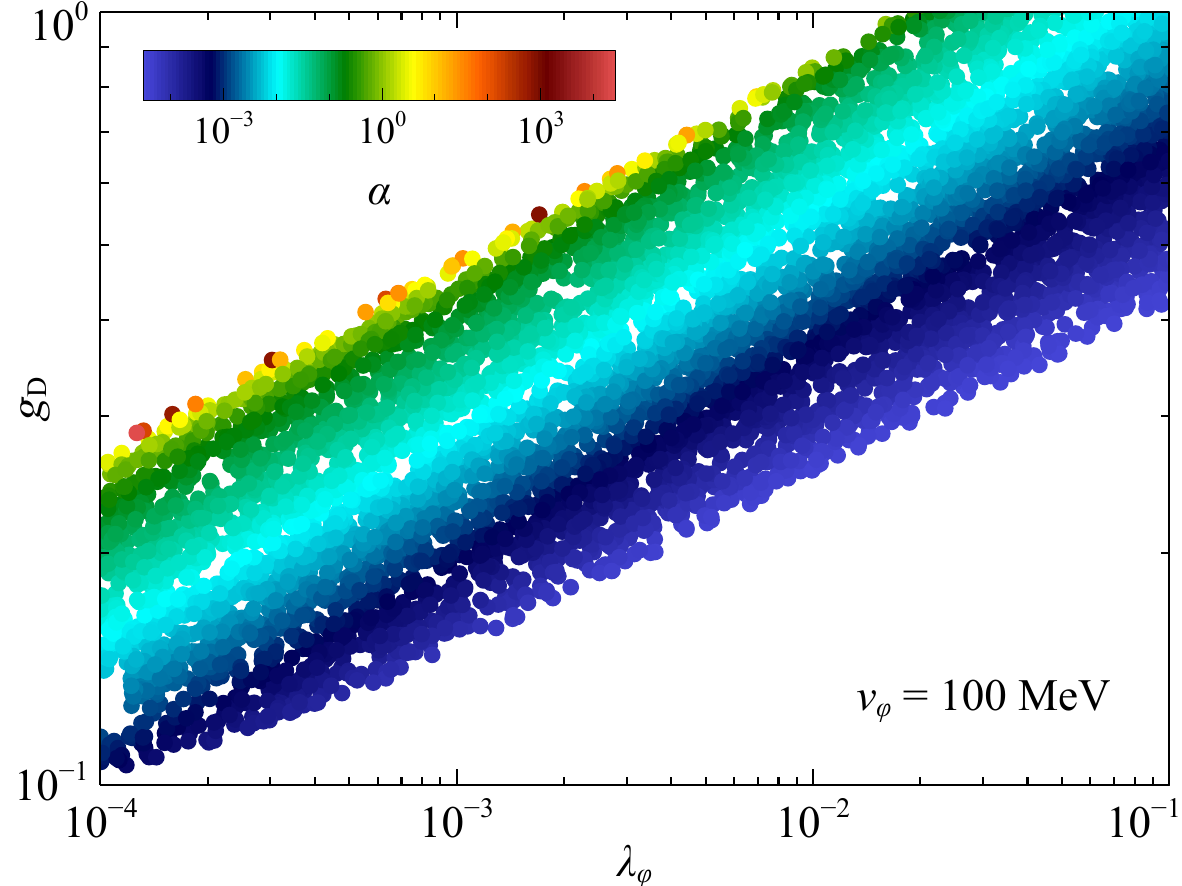}
\caption{Allowed regions in the dark sector gauge coupling ($\textsl{g}_D$) vs scalar self-coupling ($\lambda_\phi$) plane for a fixed scalar VEV $v_\phi = 100\,$MeV with the color scale representing the strength of the first-order phase transition (quantified by $\alpha$), with larger values indicating stronger transitions.}
\label{fig:gdVSlphi}
\end{figure}
In Fig.~\ref{fig:gdVSlphi} we show the region of parameter space in the $\textsl{g}_D$–$\lambda_\phi$ plane that leads to a first-order phase transition for a fixed dark scalar VEV $v_\phi = 100~{\rm MeV}$. The color code indicates the corresponding value of $\alpha(T_n)$, providing a measure of the transition strength. This plot highlights the range of gauge and scalar couplings for which the dark sector undergoes a sufficiently strong FOPT to generate potentially detectable gravitational waves.

%%%%%%%%%%%%%%%%%%%%%%%%%%%%%%%%%%%%%%%%%%%%%%%%%%%%%%%%%%%%%%%%%%%%%%%%%%%%%%%%%%%%%%%%%%%%%%%%%%%%%%%%
\section{Forbidden dark matter from first-order phase transition}\label{sec:fdmfopt}
\noindent
Having identified regions of the dark sector parameter space where the $U(1)_D$ symmetry breaking proceeds via a strong first-order phase transition, we now turn to the DM phenomenology of the same setup. In particular, we analyze how the phase transition reshapes the available annihilation channels and enables the forbidden process to control the relic abundance of $\chi$.

\begin{figure}[h]
\centering
\includegraphics[width=0.235\textwidth]{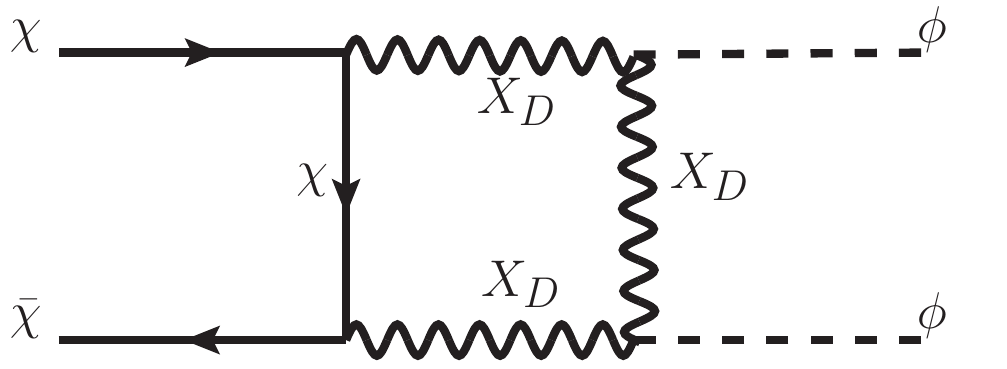}
\includegraphics[width=0.235\textwidth]{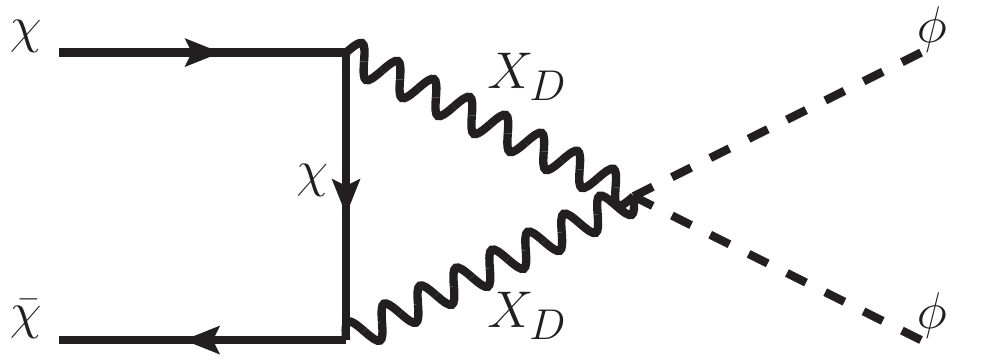}
\caption{Feynman diagrams for $\bar{\chi}\chi\rightarrow\phi\phi$ channel arising at one loop.}
\label{fig:loopFD}
\end{figure}

Before detailing the cross-sections, it is crucial to explicitly define the kinematically and dynamically allowed annihilation channels dictated by the $U(1)_D$ gauge structure. Given the gauge charge assignments of the dark matter ($Q_\chi = +1$) and the dark scalar ($Q_\Phi = -1$), a Yukawa interaction of the form $\bar{\chi} \Phi \chi$ is forbidden in the Lagrangian by gauge invariance; consequently, no tree-level $\bar{\chi}\chi\phi$ vertex exists. Furthermore, expanding the scalar kinetic term $|D_\mu \Phi|^2$ after spontaneous symmetry breaking generates a dimensionful $X_D X_D \phi$ vertex proportional to $\textsl{g}_D^2 v_\phi$, but does not yield an $X_D \phi \phi$ vertex for the real physical scalar $\phi$. In the simultaneous absence of $\bar{\chi}\chi\phi$ and $X_D \phi \phi$ vertices, the process $\chi\bar{\chi} \rightarrow \phi\phi$ cannot physically occur at tree level, even within the parameter space where $M_\chi > M_\phi$. While the $\chi\bar{\chi} \rightarrow \phi\phi$ annihilation is strictly absent at tree level, it is generated at the one-loop level via box and triangle diagrams involving internal dark fermion and dark gauge boson as shown in the Fig. \ref{fig:loopFD}.
The effective interaction generated by integrating out the gauge bosons takes the form
\begin{eqnarray}
    \mathcal{L}_{\rm eff}\simeq\frac{12\textsl{g}_D^4}{16\pi^2}\frac{M_\chi}{M_{X_D}^2}\bar{\chi}\chi\phi^2.
\end{eqnarray}
Since parity conservation forbids the $s$-wave annihilation of a fermion pair into identical scalars, this loop-induced $\phi\phi$ channel strictly proceeds via $p$-wave and suffers a velocity suppression ($v^2$) at freeze-out and later times.

Consequently, in our setup, four annihilation channels can in principle contribute to setting the DM abundance: \begin{eqnarray}
\bar{\chi}\chi \rightarrow X_D X_D, ~
\bar{\chi}\chi \rightarrow X_D \phi, ~
\bar{\chi}\chi \rightarrow \phi\phi, ~
\bar{\chi}\chi \rightarrow \bar{f} f
\end{eqnarray}
where $f$ denotes SM fermions. The process $\bar{\chi}\chi \rightarrow \bar{f} f$ is induced by the kinetic mixing between the dark gauge boson $X_D$ and the hypercharge gauge boson, but its contribution to the relic density can be made negligible by taking the kinetic mixing parameter $\epsilon$ sufficiently small. 
In this regime, the DM abundance is primarily determined by annihilations into dark-sector states.

As discussed in Sec. \ref{sec:fopt}, and evident from Fig.~\ref{fig:gdVSlphi}, a strong first-order phase transition in the dark sector dictates $\textsl{g}_D \gg \lambda_\Phi$, which naturally enforces a strict mass hierarchy $M_{X_D} \gg M_\phi$ at fixed $v_\phi$. Consequently, the phase-space suppression for $\bar{\chi}\chi \rightarrow X_D X_D$ is more severe than for the mixed final state $\bar{\chi}\chi \rightarrow X_D \phi$. Thus, the tree-level, $s$-wave process $\bar{\chi}\chi \rightarrow X_D \phi$ typically dominates the annihilation rate in the broken phase. To realize the forbidden DM mechanism, we restrict our parameter space such that this dominant channel is always kinematically forbidden at zero temperature:
\begin{eqnarray}
2 M_\chi < M_{X_D} + M_\phi.
\end{eqnarray}
This channel thus becomes active in the early Universe solely due to the high-energy Boltzmann tail of the thermal distribution. While $X_D \phi$ remains universally forbidden in our analysis, the loop-induced $\bar{\chi}\chi \rightarrow \phi\phi$ channel can be either allowed ($M_\chi > M_\phi$) or forbidden ($M_\chi < M_\phi$) depending on the specific scalar mass. In the following subsections, we analyze these two distinct phenomenological regimes and delineate the parameter spaces where the relic density is governed by the intricate interplay between tree-level exponential suppression and loop-level $p$-wave suppression.

%%%%%%%%%%%%%%%%%%%%%%%%%%%%%%%%%%%%%%%%%%%%%%%%%%%%%%%%%%%%%%%%%%%%%%%%%%%%%%%%%%%%%%%%%%%%%%%%%%%%%%%%
\subsection{Dark matter relic when $\bar{\chi}\chi\rightarrow X_D\phi$ is forbidden, but $\bar{\chi}\chi\rightarrow \phi\phi$ is allowed}
\noindent
We first discuss the scenario where the loop-induced $\bar{\chi}\chi\rightarrow \phi\phi$ process is kinematically allowed while the tree-level $\bar{\chi}\chi\rightarrow X_D\phi$ channel remains strictly forbidden. We define the mass splitting parameter as $\Delta_{X_D\phi}=\frac{M_{X_D}+M_\phi-2M_\chi}{2M_\chi}$. The DM yield $Y_\chi \equiv n_\chi/s$ evolves according to the Boltzmann equation:
\begin{eqnarray}
\frac{dY_\chi}{dz}&=&-\frac{sz}{\mathcal{H}(M_{\chi})}
\left(\langle\sigma v\rangle^f_{\bar{\chi}\chi\rightarrow X_D\phi}+\langle\sigma v\rangle^a_{\bar{\chi}\chi\rightarrow \phi\phi}\right)\nonumber\\&&\times\left(Y_\chi^2-(Y^{eq}_{\chi})^2\right).\label{eq:be_ff}
\end{eqnarray}
where $z=M_{\chi}/T$ is the dimensionless parameter used to track the evolution of temperature, $Y^{eq}_\chi$ is the equilibrium abundance of $\chi$ {\it i.e.} $Y^{eq}_\chi(z)=0.116\frac{g_\chi}{g_*}x^2K_2(z)$, $g_*$ is the effective number of relativistic d.o.f at temperature $T$, and $K_2(z)$ is the modified Bessel function of the second kind. The Hubble parameter and entropy density are respectively given by
$
\mathcal{H}(M_\chi) = 1.66\,\sqrt{g_*}\,\frac{M_\chi^2}{M_{\rm Pl}}\,,
s = \frac{2\pi^2}{45}\,g_*\,M_\chi^3 z^{-3},
$
where $M_{\rm Pl} = 1.22\times 10^{19}\,\text{GeV}$ is the Planck mass. The thermally averaged cross-section for the forbidden channel is related to the allowed reverse process via detailed balance as: 
\begin{eqnarray}
\langle\sigma v\rangle^f_{\chi \bar{\chi} \rightarrow X_D \phi}=\frac{n^{\rm eq}_{X_D}n^{\rm eq}_{\phi}}{(n^{\rm eq}_{\chi})^2} \langle\sigma v\rangle^a_{ X_D \phi\rightarrow \chi \bar{\chi}}.
\end{eqnarray}
\begin{table*}[tbh]
\centering
\caption{Benchmark points for satisfying correct DM relic via $\bar{\chi}\chi\rightarrow X_D\phi (f)$, $\bar{\chi}\chi\rightarrow \phi\phi (a)$  channels, and giving first-order phase transition. $\Delta_{X_D\phi}({BP0})=0.4091$, $\Delta_{X_D\phi}({BP1})=0.60842$, $\Delta_{X_D\phi}({BP2})=1.53066$, $\Delta_{X_D\phi}({BP3})=0.2544036$, $\Delta_{X_D\phi}({BP4})=0.411547441420$.}
    \resizebox{\textwidth}{!}{
        \begin{tblr}{
                colspec={|l|l|l|l|l|l|l|l|l|l|l|}            }
            \toprule \textit{BPs}&$M_{\chi}(\rm MeV)$&$M_{X_D}(\rm MeV)$&$M_{\phi}(\rm MeV)$& $\lambda_\Phi$&$v_\phi(\rm MeV)$&$\textsl{g}_D$&$T_c(\rm MeV)$&$T_n(\rm MeV)$ &$\alpha_*$ & $\beta_*/\mathcal{H}_*$\\
            \toprule
            \textit{BP0}&
            $42.326$& $112.059$& $7.222$&$0.000104$&$500$&$0.2241$&$52.0094$&$35.432$&$0.1332$&$8880.491$\\\hline
            \textit{BP1}&
            $23.1786$& $54.6895$& $5.840255$&$0.0017054$&$100$&$0.546895$&$17.41913$&$1.516388$&$839.4085$&$334.49253$\\\hline
            \textit{BP2}&
            $23.4358$& $98.49544$& $20.120645$&$0.02024$&$100$&$0.984954$&$35.27115$&$24.25950$&$0.16288$&$658.2282$\\\hline
            \textit{BP3}&
            $22.2034$& $49.60992$& $6.094203$&$0.00185696$&$100$&$0.496099$&$20.237876$&$13.308489$&$0.204574$&$1809.8969$\\\hline
            \textit{BP4}&
            $26.919842$& $49.387419$& $26.60985$&$0.03540420$&$100$&$0.493874198$&$75.08041$&$74.9224079$&$0.00041969$&$139918.089$\\\hline
            \textit{BP5}&
            $4778.9589$& $10713.3738$& $1164.1505$&$0.001694$&$20000$&$0.5356686$&$3564.6272$&$1143.8917$&$0.86177$&$849.7935$\\\hline
            \textit{BP6}&
            $7912.7921$& $19649.45091$& $3751.20211$&$0.017589$&$20000$&$0.98247254$&$6535.093$&$3046.63690$&$0.152777$&$191.2885$\\
            \bottomrule
    \end{tblr}}
    \label{tab:tab1fa}
\end{table*}
\begin{figure*}[tbh]
\centering
\includegraphics[width=0.495\textwidth]{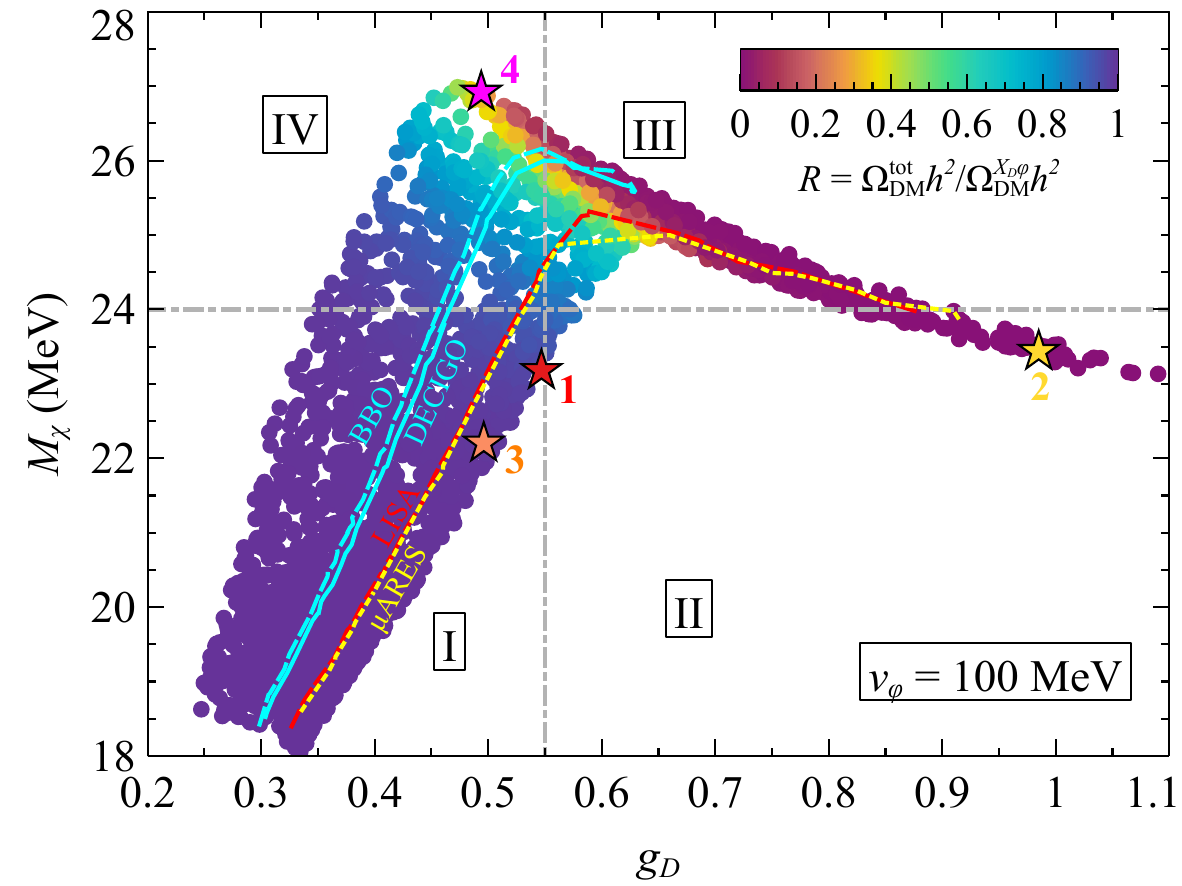}
\includegraphics[width=0.495\textwidth]{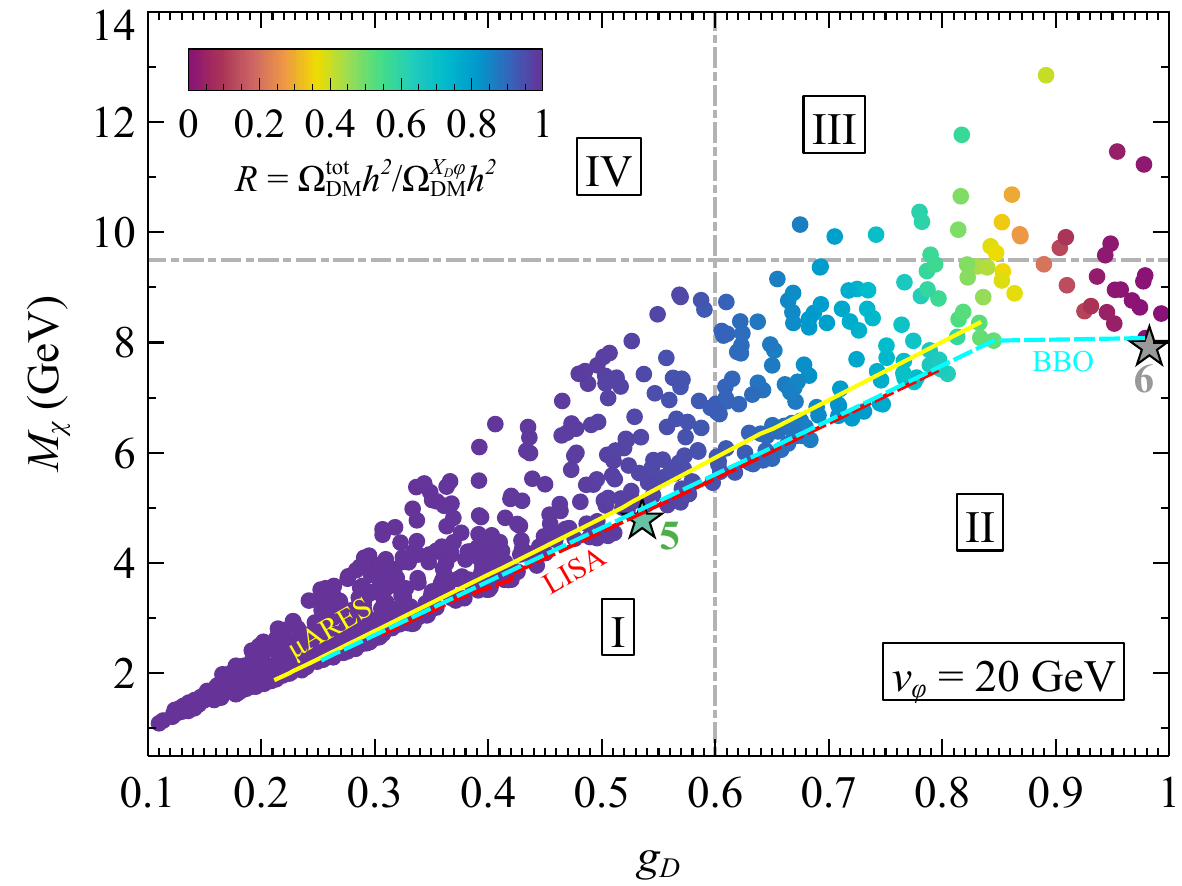}
\caption{[\textit{Left}:] correct DM relic parameter space in the plane of $M_\chi$ vs $\textsl{g}_D$ for fixed VEV of $\phi$, $v_\phi=100$ MeV. The same points also give rise to a first-order phase transition. The color code represents the ratio $R$. The SNR contours are shown for BBO \cite{Yagi:2011wg} (cyan dashed), DECIGO \cite{Seto:2001qf} (cyan solid), LISA \cite{LISA:2017pwj} (red solid), and $\mu$ARES \cite{Sesana:2019vho} (yellow dotted) for $\mathcal{S}=10$ with $\tau_{\rm obs}=1$ year. See section \ref{sec:gwfopt} for more details on SNR. [\textit{Right}:] the same with $v_\phi=20$ GeV.} 
\label{fig:relicpar_scan1}
\end{figure*}

Solving the Boltzmann equation yields the late-time value of $Y_\chi$, from which the present relic abundance follows as
\begin{eqnarray}
\Omega_{\rm DM}h^2\simeq0.118\left(\frac{Y_\chi}{4.2\times10^{-10}}\right)\left(\frac{M_\chi}{1~\rm GeV}\right).
\end{eqnarray}

\begin{figure*}[tbh]
\centering
\includegraphics[width=0.495\textwidth]{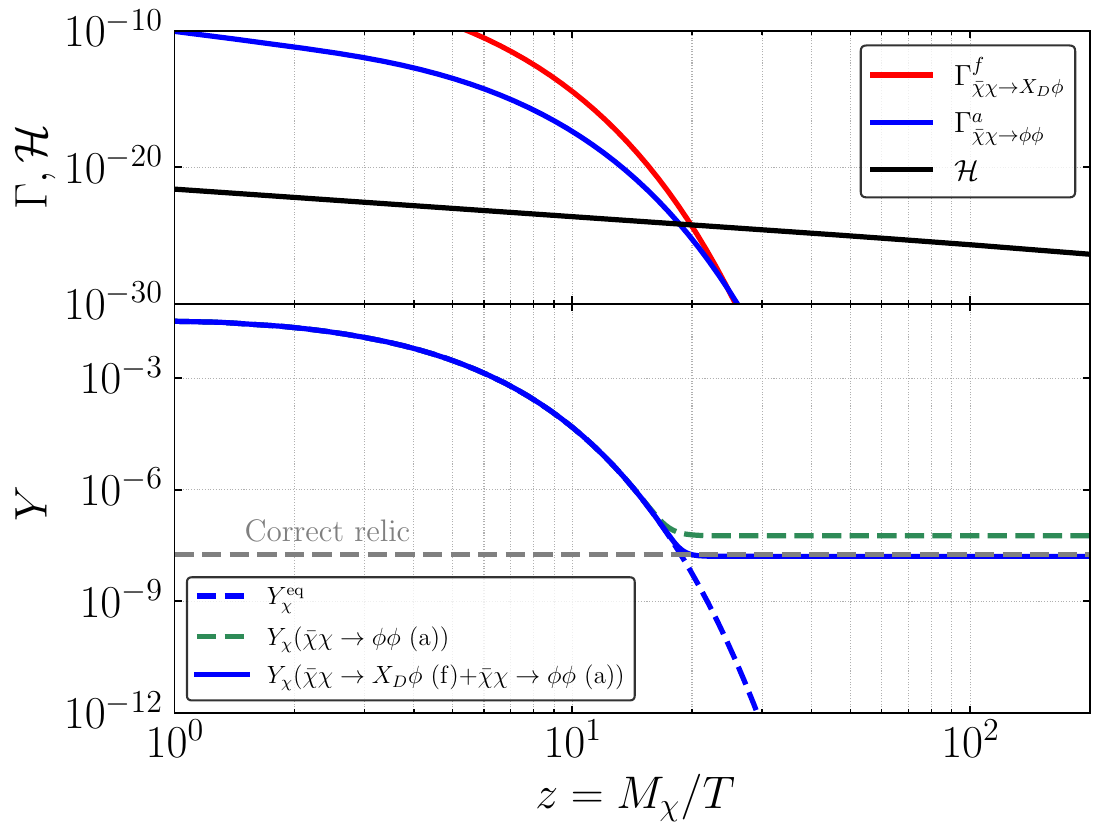}
\includegraphics[width=0.495\textwidth]{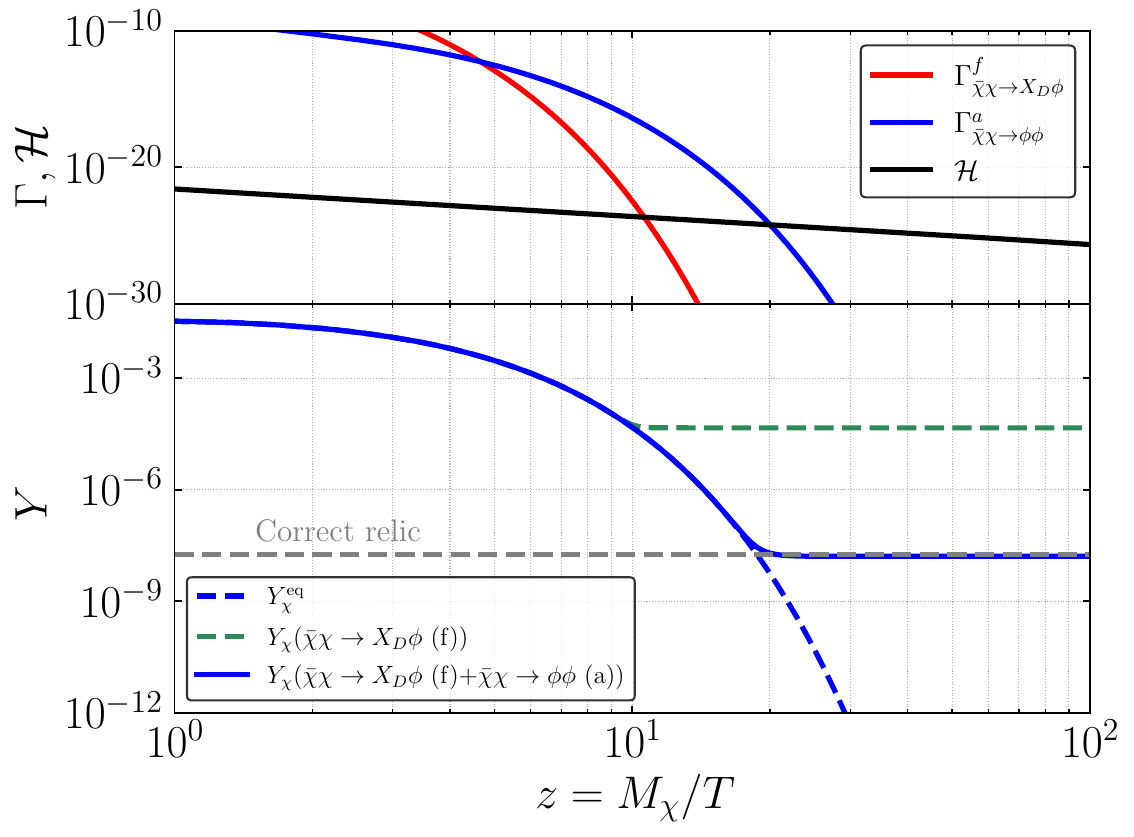}
\caption{[\textit{Left}:] interaction rates comparison with Hubble (\textit{top panel}) and cosmological evolution of the DM (\textit{bottom panel}) for BP1 from Table \ref{tab:tab1fa}. [\textit{Right}:] same for BP2 from Table \ref{tab:tab1fa}.}
\label{fig:fabp12}
\end{figure*}

\begin{figure*}[tbh]
\centering
\includegraphics[width=0.495\textwidth]{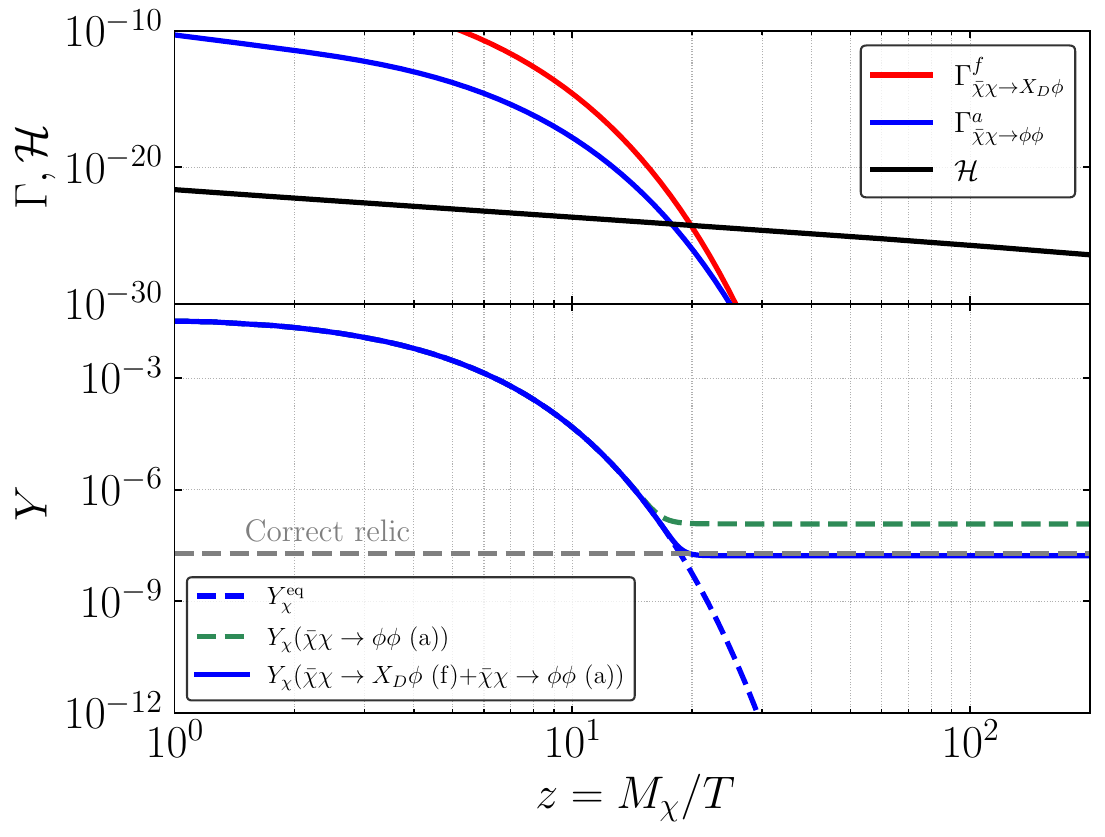}
\includegraphics[width=0.495\textwidth]{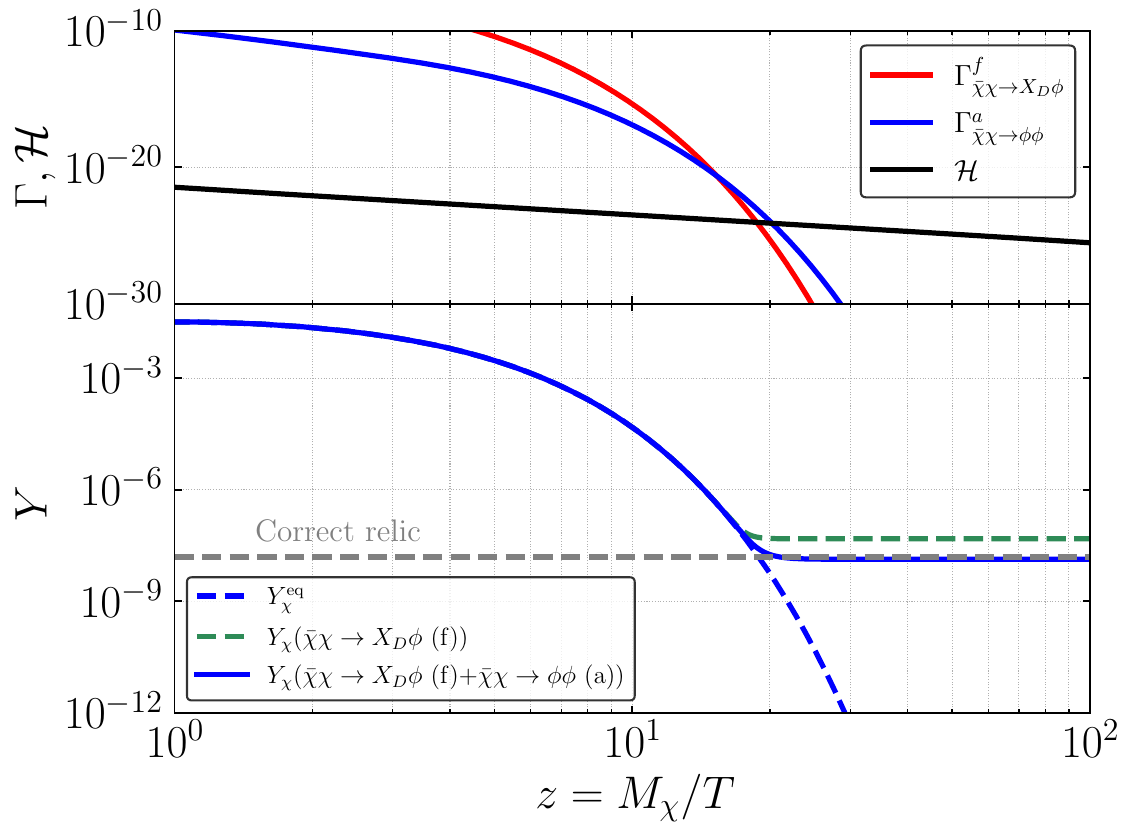}
\caption{[\textit{Left}:] interaction rates comparison with Hubble (\textit{top panel}) and cosmological evolution of the DM (\textit{bottom panel}) for BP3 from Table \ref{tab:tab1fa}. [\textit{Right}:] same for BP4 from Table \ref{tab:tab1fa}.}
\label{fig:fabp34}
\end{figure*}

In the \textit{left} panel of Fig. \ref{fig:relicpar_scan1}, we map the correct relic parameter space in the $M_\chi-\textsl{g}_D$ plane for a fixed scalar VEV: $v_\phi=100$ MeV. The color code represents the ratio
\begin{eqnarray}
R=\Omega_{\rm DM}^{\rm tot}h^2/\Omega_{\rm DM}^{X_D\phi}h^2,
\end{eqnarray}
where $\Omega_{\rm DM}^{\rm tot}h^2$ is obtained by solving Eq \ref{eq:be_ff} while $\Omega_{\rm DM}^{X_D\phi}h^2$ is obtain from the same Eq \ref{eq:be_ff} but keeping $\bar{\chi}\chi\rightarrow\phi\phi$ channel switched off. We note that $R$ quantifies the relative importance of the two channels. $R$ approaching unity indicates that the heavily Boltzmann-suppressed, yet strongly-coupled $s$-wave $\bar{\chi}\chi \rightarrow X_D\phi$ channel dictates the relic density. Conversely, $R \ll 1$ indicates that the exponential penalty on $\bar{\chi}\chi \rightarrow X_D\phi$ is too severe, allowing the phase-space open (but loop and $p$-wave suppressed) $\bar{\chi}\chi \rightarrow \phi\phi$ channel to take over. As the DM mass $M_\chi$ increases, the cross-section for the allowed $\bar{\chi}\chi \rightarrow \phi\phi$ channel inherently grows due to the mass insertion factor $(M_\chi/M^2_{X_D})^2$ embedded in its effective coupling. At the same time, maintaining the correct relic density and FOPT conditions across the parameter space often requires a larger scalar mass $M_\phi$. This larger $M_\phi$ increases the required mass splitting $\Delta_{X_D\phi}$, which inflicts a severe exponential penalty on the $\bar{\chi}\chi \rightarrow X_D\phi$ channel until the $\bar{\chi}\chi \rightarrow \phi\phi$ process inevitably dominates.

To illustrate this dynamical crossover, we select four benchmark points (BP1–BP4) listed in Table \ref{tab:tab1fa}. Comparing BP1 (red star) and BP2 (yellow star), we nearly fix the DM mass but significantly increase $M_{X_D}$ (and thus $\textsl{g}_D$). In the \textit{left} panel of Fig. \ref{fig:fabp12} (BP1), the allowed $\bar{\chi}\chi \rightarrow \phi\phi$ channel (blue curve) freezes out early due to its loop suppression, while the forbidden $\bar{\chi}\chi \rightarrow X_D\phi$ channel (red curve) remains active longer, solely determining the relic abundance. However, for BP2 (\textit{right} panel), the increased $M_{X_D}$ leads to a much larger mass splitting ($\Delta_{X_D\phi}({BP2}) \approx 1.53$ vs $\Delta_{X_D\phi}({BP1}) \approx 0.61$). This inflicts a massive exponential penalty on the $\bar{\chi}\chi \rightarrow X_D\phi$ channel, causing it to decouple early. Consequently, the relic abundance for BP2 is entirely governed by the $\textsl{g}_D^8$-enhanced $\bar{\chi}\chi \rightarrow \phi\phi$ process. Conversely, examining BP3 (orange star) and BP4 (magenta star) in Fig.~\ref{fig:fabp34}, we keep $M_{X_D}$ nearly fixed while increasing $M_\chi$. In BP3, $\bar{\chi}\chi \rightarrow X_D\phi$ dominates. For BP4, the simultaneous requirement of the correct relic density dictates a notably heavier dark scalar ($M_\phi \approx 26.6$ MeV compared to $6.1$ MeV in BP3). This larger scalar mass drives up the mass-splitting parameter ($\Delta_{X_D\phi}({BP4}) \approx 0.41$ compared to $\Delta_{X_D\phi}({BP3}) \approx 0.25$), heavily suppressing the $\bar{\chi}\chi \rightarrow X_D\phi$ channel. Concurrently, the larger $M_\chi$ boosts the $(M_\chi/M^2_{X_D})^2$ mass-insertion factor, causing the $\bar{\chi}\chi \rightarrow \phi\phi$ channel to definitively govern the freeze-out. This same qualitative interplay is observed for a higher symmetry-breaking scale ($v_\phi = 20$ GeV) in the \textit{right} panel of Fig.~\ref{fig:relicpar_scan1}, represented by BP5 and BP6.

To explicitly connect this particle physics parameter space with upcoming macroscopic cosmological probes, we have additionally overlaid Gravitational Wave Signal-to-Noise Ratio (SNR) ($\mathcal{S}$) contours for various future observatories on Fig.~\ref{fig:relicpar_scan1}. The contours represent a detection threshold of $\mathcal{S}=10$ with an observation time of $\tau_{\rm obs}=1$ year for BBO \cite{Yagi:2011wg} (cyan dashed), DECIGO \cite{Seto:2001qf} (cyan solid), LISA \cite{LISA:2017pwj} (red solid), and $\mu$ARES \cite{Sesana:2019vho} (yellow dotted). These contours visually demonstrate the gravitational wave detectability prospects directly alongside the valid dark matter relic parameter space; a detailed discussion of the exact SNR calculation and the corresponding phase transition dynamics is dedicated to Section~\ref{sec:gwfopt}.

\begin{table*}[tbh]
    \centering
    \caption{Benchmark points for satisfying correct DM relic via $\bar{\chi}\chi\rightarrow X_D\phi (f)$, $\bar{\chi}\chi\rightarrow \phi\phi (f)$  channels, and giving first-order phase transition. $\Delta_{\phi\phi}(BP7)=0.21$, $\Delta_{\phi\phi}(BP8)=0.097139$, $\Delta_{\phi\phi}(BP9)=0.0947368$, $\Delta_{\phi\phi}(BP10)=0.0138817$.}
    \resizebox{\textwidth}{!}{
    \begin{tblr}{
                colspec={|l|l|l|l|l|l|l|l|l|l|l|}            }
            \toprule \textit{BPs}&$M_{\chi}(\rm MeV)$&$M_{X_D}(\rm MeV)$&$M_{\phi}(\rm MeV)$& $\lambda_\Phi$&$v_\phi(\rm MeV)$&$\textsl{g}_D$&$T_c(\rm MeV)$&$T_n(\rm MeV)$ &$\alpha_*$ & $\beta_*/\mathcal{H}_*$\\
            \toprule
            \textit{BP7}&
            $30.53$& $145.41$& $36.96$&$0.06829$&$100$&$1.4541$&$47.3661$&$24.4982$&$0.39091$&$95.04275$\\\hline
            \textit{BP8}&
            $81$& $479.225$& $88.869$&$0.015796$&$500$&$0.95845$&$158.0208$&$71.6957$&$0.7169459$&$187.9065$\\\hline
            \textit{BP9}&
            $257.181$& $1226.57$& $281.545$&$0.039634$&$1000$&$1.22657$&$408.497$&$241.9205$&$0.086404$&$269.9877$\\\hline
            \textit{BP10}&
            $1768.0408$& $6761.662$& $1792.5843$&$0.06426717$&$5000$&$1.35233$&$2453.449$&$1869.49$&$0.0186588$&$704.74155$\\
            \bottomrule
    \end{tblr}}
    \label{tab:tab1ff}
\end{table*}
\begin{figure*}[tbh]
\centering
\includegraphics[width=0.495\textwidth]{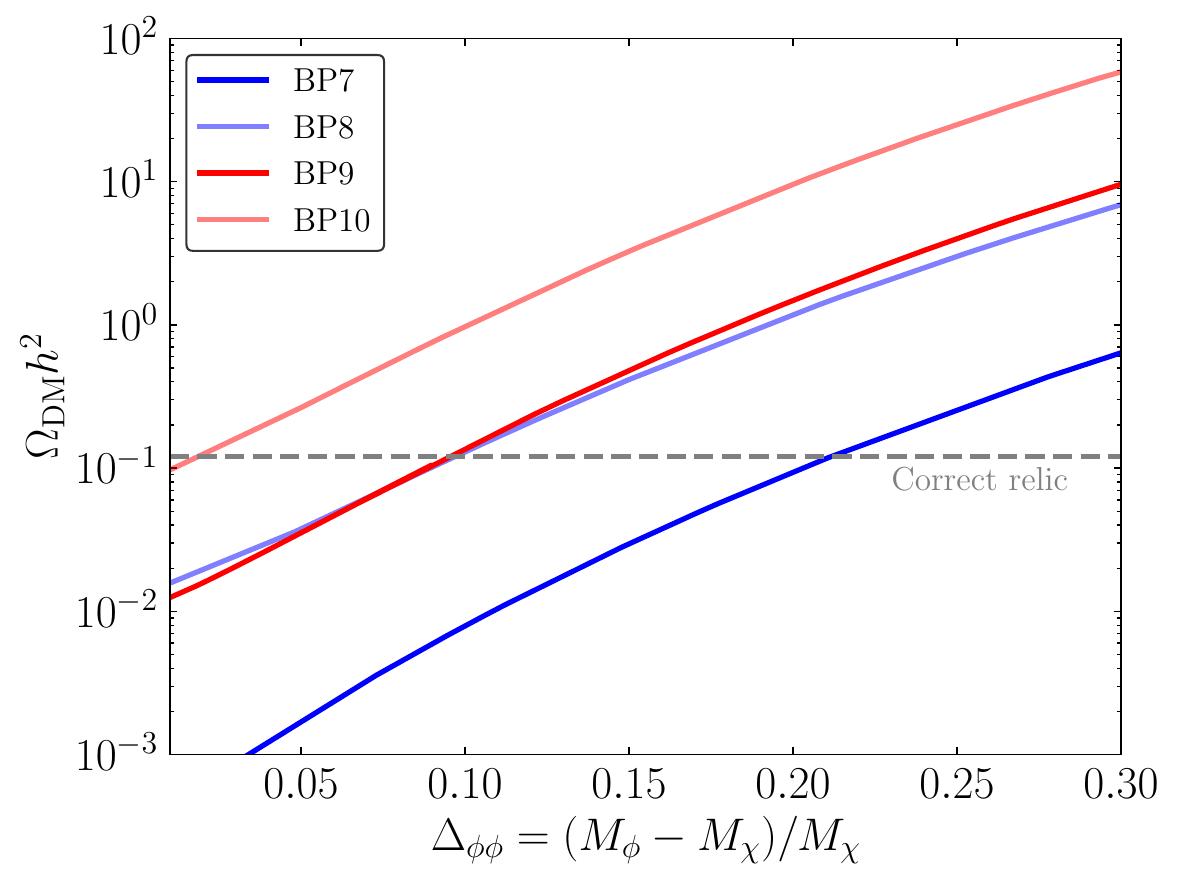}
\includegraphics[width=0.495\textwidth]{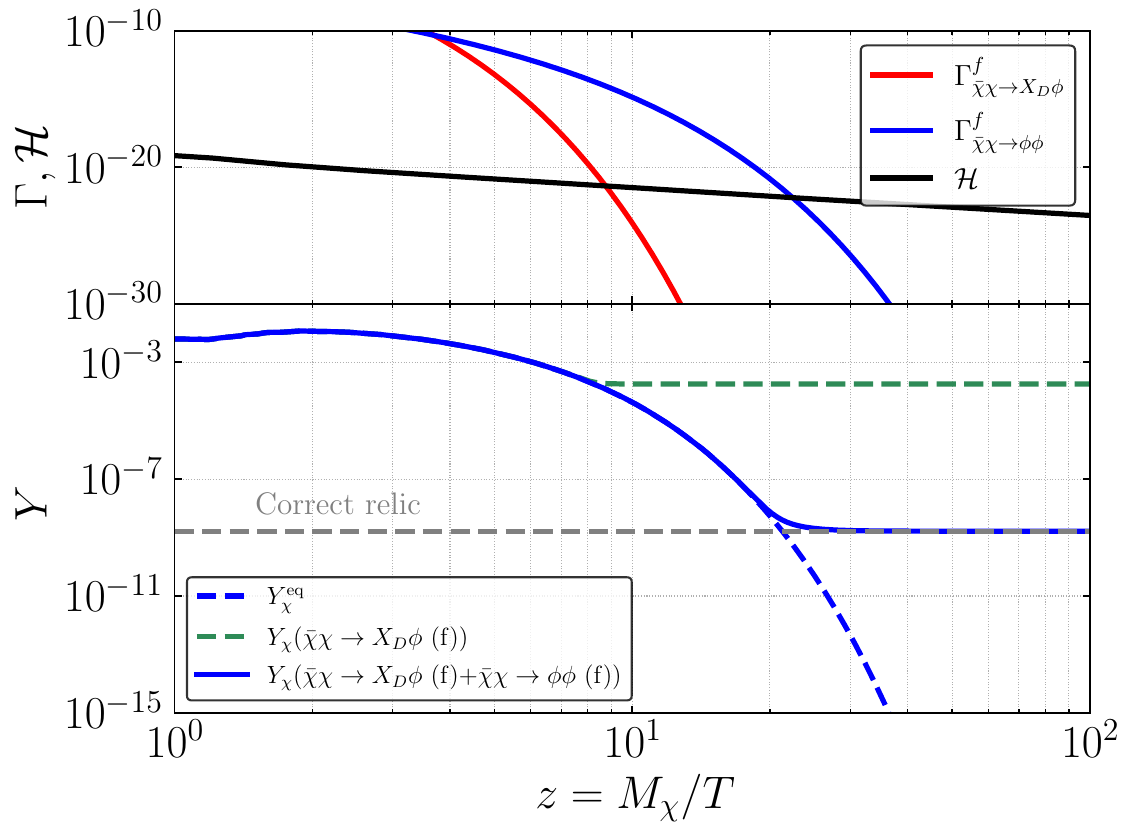}
\caption{[\textit{Left}:] dark matter relic density $\Omega_{\rm DM}h^2$ as a function of $\Delta_{\phi\phi}$ for the benchmark points in Table \ref{tab:tab1ff}. [\textit{Right}:] interaction rates comparison with Hubble (\textit{top panel}) and cosmological evolution of the DM (\textit{bottom panel}) for BP9 from Table \ref{tab:tab1ff}.}
\label{fig:relic_case2}
\end{figure*}

%%%%%%%%%%%%%%%%%%%%%%%%%%%%%%%%%%%%%%%%%%%%%%%%%%%%%%%%%%%%%%%%%%%%%%%%%%%%%%%%%%%%%%%%%%%%%%%%%%%%%%%%
\subsection{Dark matter relic when $\bar{\chi}\chi\rightarrow X_D\phi$, and $\bar{\chi}\chi\rightarrow \phi\phi$ both are forbidden}
\noindent
We now examine the regime where $M_\chi < M_\phi$, rendering both the $\bar{\chi}\chi \rightarrow X_D\phi$ and the loop-induced $\bar{\chi}\chi \rightarrow \phi\phi$ channels kinematically forbidden at zero temperature. We define the mass splitting for the $\bar{\chi}\chi \rightarrow \phi\phi$ channel as $\Delta_{\phi\phi}=\frac{M_\phi-M_\chi}{M_\chi}$. The relic density is obtained by solving the Boltzmann equation:
\begin{eqnarray}
\frac{dY_\chi}{dz}&=&-\frac{sz}{\mathcal{H}(M_{\chi})}
\left(\langle\sigma v\rangle^f_{\bar{\chi}\chi\rightarrow X_D\phi}+\langle\sigma v\rangle^f_{\bar{\chi}\chi\rightarrow \phi\phi}\right)\nonumber\\&&\times\left(Y_\chi^2-(Y^{eq}_{\chi})^2\right),
\end{eqnarray}
where both forbidden cross-sections are detailed-balanced from their respective allowed processes:
\begin{eqnarray}
\langle\sigma v\rangle^f_{\chi \bar{\chi} \rightarrow X_D \phi}&=&\frac{n^{\rm eq}_{X_D}n^{\rm eq}_{\phi}}{(n^{\rm eq}_{\chi})^2} \langle\sigma v\rangle^a_{ X_D \phi\rightarrow \chi \bar{\chi}},\nonumber\\
\langle\sigma v\rangle^f_{\chi \bar{\chi} \rightarrow \phi \phi}&=&\frac{(n^{\rm eq}_{\phi})^2}{(n^{\rm eq}_{\chi})^2} \langle\sigma v\rangle^a_{ \phi \phi\rightarrow \chi \bar{\chi}}.
\end{eqnarray}

\begin{figure*}[tbh]
\centering
\includegraphics[width=0.495\textwidth]{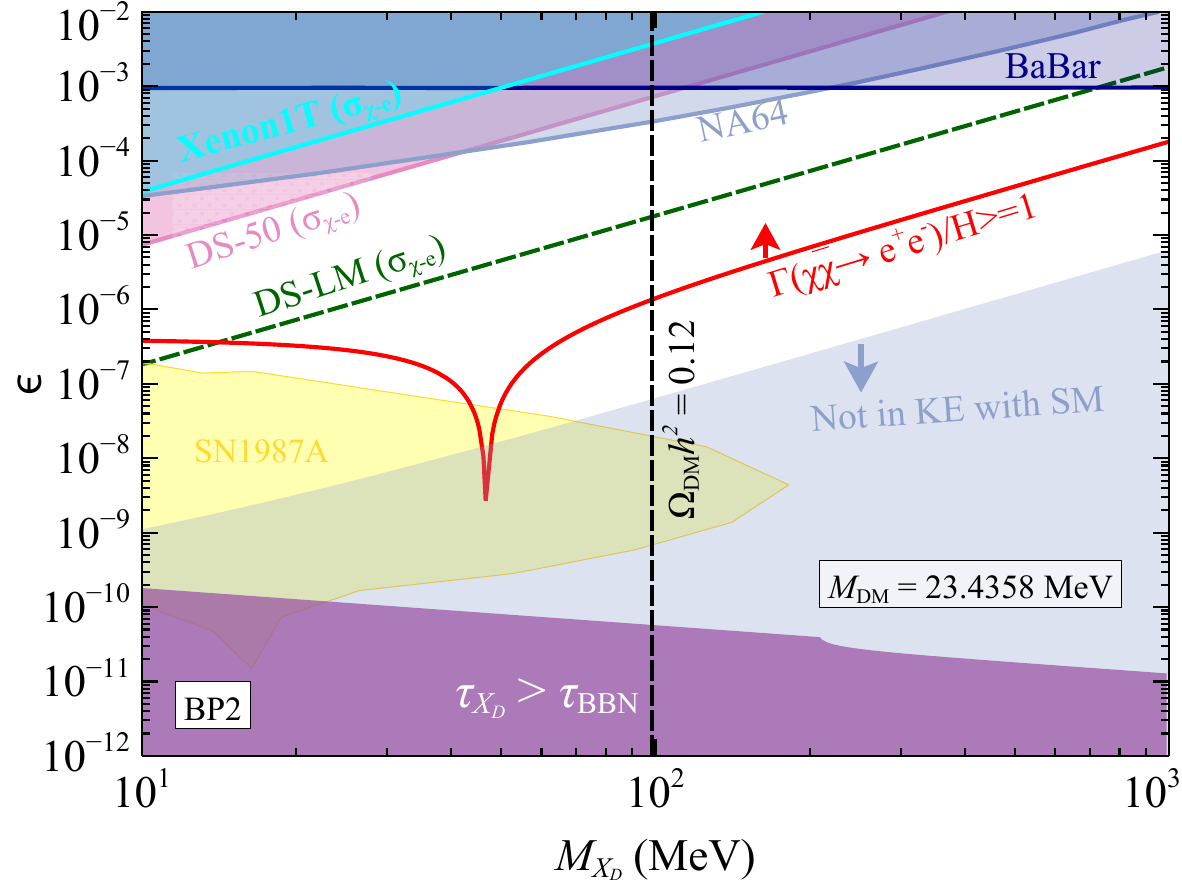}
\includegraphics[width=0.495\textwidth]{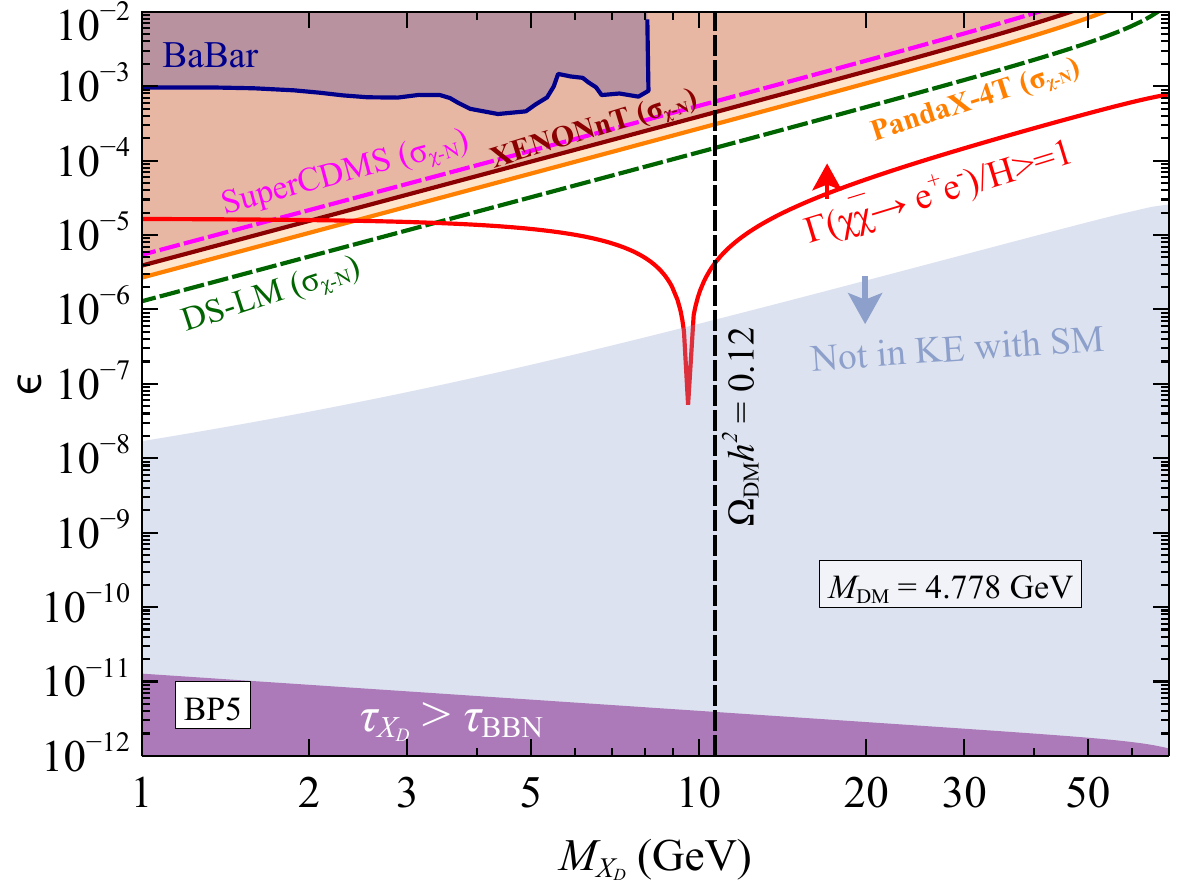}\\
\includegraphics[width=0.495\textwidth]{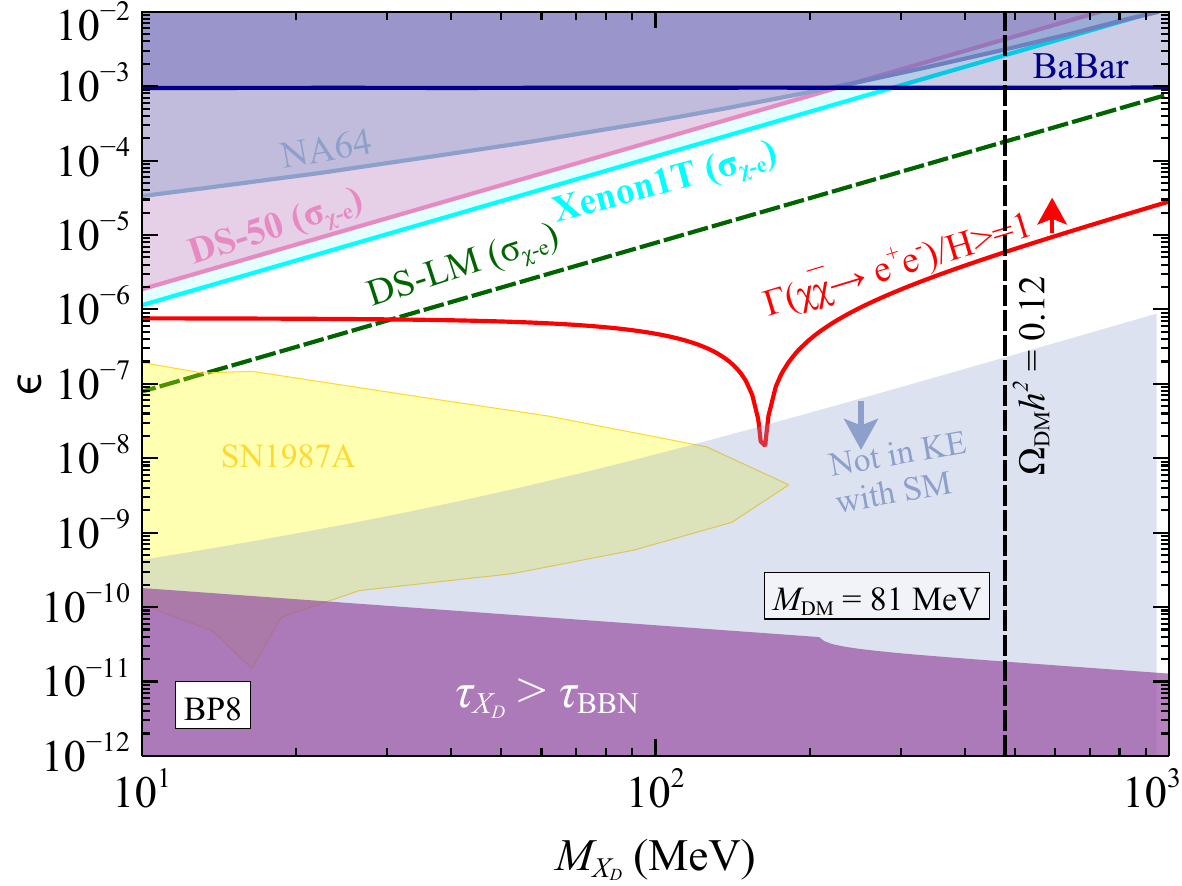}
\includegraphics[width=0.495\textwidth]{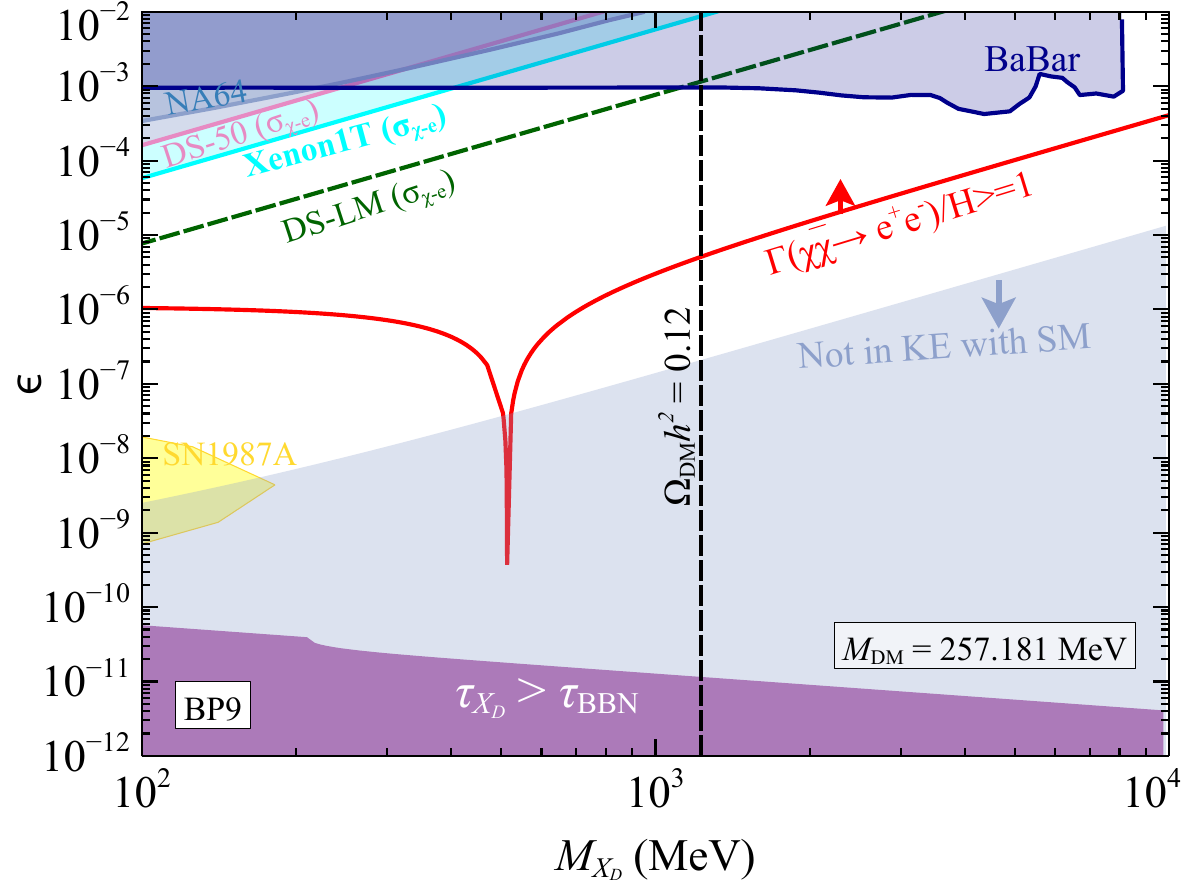}
\caption{Parameter space in the $\epsilon$–$M_{X_D}$ plane for \textit{BP2}, \textit{BP5}, \textit{BP8}, and \textit{BP9} (top left, top right, bottom left, bottom right respectively), as defined in Tables~\ref{tab:tab1fa} and \ref{tab:tab1ff}. Shaded regions indicate possible constraints; the vertical black dashed line shows the value of $M_{X_D}$ that yields the correct dark matter relic abundance for each benchmark.}
\label{fig:ddparameterspace}
\end{figure*}

In this completely forbidden regime, the relic abundance depends on a competition between distinct suppressions. Because the FOPT demands $M_{X_D} \gg M_\phi$, the kinematic threshold for the $\bar{\chi}\chi \rightarrow X_D\phi$ channel is much larger than the threshold for the $\bar{\chi}\chi \rightarrow \phi\phi$ channel. Thus, the $\bar{\chi}\chi \rightarrow X_D\phi$ channel suffers a significantly larger exponential Boltzmann penalty compared to the milder penalty on the $\bar{\chi}\chi \rightarrow \phi\phi$ channel. When this disparity in Boltzmann suppression overcomes the intrinsic loop and $p$-wave penalties of the $\bar{\chi}\chi \rightarrow \phi\phi$ process, the $\bar{\chi}\chi \rightarrow \phi\phi$ channel dominates the freeze-out. In the \textit{left} panel of Fig.~\ref{fig:relic_case2}, we show the DM relic abundance as a function of $\Delta_{\phi\phi}$ for BP7--BP10. The \textit{right} panel explicitly illustrates this dynamic for BP9: the milder exponential penalty allows the $\bar{\chi}\chi \rightarrow \phi\phi$ channel to remain in equilibrium longer, cleanly determining the final DM relic abundance.

Finally, in Fig.~\ref{fig:ddparameterspace}, we present the viable parameter space in the $\epsilon$–$M_{X_D}$ plane for selected benchmarks across both regimes i.e. \textit{BP2}, \textit{BP5}, \textit{BP8}, and \textit{BP9} (top left, top right, bottom left, and bottom right, respectively), as defined in Tables~\ref{tab:tab1fa} and \ref{tab:tab1ff}. For each benchmark, the shaded and hatched regions summarize current constraints from DM-electron and DM–nucleon scattering experiments, as well as bounds from accelerator and astrophysical probes, while the vertical black dashed line indicates the value of $M_{X_D}$ that reproduces the observed relic abundance. The light blue region corresponds to parameter choices where $\chi$ fails to maintain kinetic equilibrium with the SM bath, primarily determined by the $t$-channel process $\chi e^- \to \chi e^-$. The region above the red solid curve satisfies $\Gamma_{\chi\bar{\chi}\rightarrow e^+e^-}/H \gtrsim 1$ at $z=10$, implying that annihilations into $e^+e^-$ substantially modify the relic density and the simple forbidden-channel picture no longer applies. The purple band near $\epsilon \sim 10^{-11}$ is excluded by big bang nucleosynthesis (BBN) considerations, since in this regime the dark gauge boson $X_D$ becomes too long-lived and disrupts standard nucleosynthesis.

In Fig.~\ref{fig:ddparameterspace}, current exclusion limits from DM–$e$ scattering experiments, specifically DS-50~\cite{DarkSide:2022knj} and XENON1T~\cite{XENON:2019gfn}, are depicted as pink and cyan shaded regions, respectively. Correspondingly, direct-detection exclusion limits from the DM–nucleon experiments XENONnT~\cite{XENON:2024hup} and PandaX-4T~\cite{PandaX:2025rrz} are indicated by the dark red and orange shaded regions. To highlight the future testability of the allowed parameter space, projected sensitivities from upcoming experiments are overlaid as non-shaded lines: the sensitivity of DS-LM~\cite{GlobalArgonDarkMatter:2022ppc} is represented by green dashed lines, while the projected reach of SuperCDMS~\cite{SuperCDMS:2016wui} is shown as a magenta dashed line.
Additional constraints from NA64~\cite{Banerjee:2019pds} (light blue shaded), BaBar~\cite{BaBar:2014zli} (dark blue shaded), and SN1987A~\cite{Chang:2016ntp,Caputo:2025avc,Caputo:2025aac} (light orange shaded) are included as well. Similar features and trends are visible across all four benchmark panels, illustrating that the interplay between relic-density requirements, kinetic equilibrium, and the various experimental and cosmological bounds consistently carves out a narrow but well-defined region in $\epsilon$–$M_{X_D}$ space.

%%%%%%%%%%%%%%%%%%%%%%%%%%%%%%%%%%%%%%%%%%%%%%%%%%%%%%%%%%%%%%%%%%%%%%%%%%
\section{Other relevant constraints}\label{sec:constraint}
\noindent
In addition to relic-density, direct-detection, and other requirements discussed in the previous section, the scenario is subject to several cosmological and collider bounds that further shape the viable parameter space. In this section, we discuss the limits from BBN on the lifetime of the scalar $\phi$, and bounds on invisible Higgs decays from LHC measurements.
%%%%%%%%%%%%%%%%%%%%%%%%%%%%%%%%%%%%%%%%%%%%%%%%%%%%%%%%%%%%%%%%%%%%%%%%%%
\subsection{BBN constraint from $\Phi$ decay}
\noindent
The light scalar, $\phi$, can decay into the SM fermions after mixing with the SM Higgs boson. The partial decay width into a fermion pair is given as
\begin{eqnarray}
\Gamma_{\phi\rightarrow ff}=\sin^2\theta\left(\frac{m_f}{v_0}\right)^2\frac{M_{\phi}}{8\pi}\bigg(1-\frac{4m_f^2}{M^2_{\phi}}\bigg)^{\frac{3}{2}},
\end{eqnarray}
where $m_f$ is the mass of the fermion, $\sin\theta$ is the mixing angle between the SM Higgs and scalar $\phi$, $v_0$ is the VEV of SM Higgs.
The corresponding lifetime is given by
\begin{eqnarray}
\tau_{\phi}=(\Gamma_{\phi\rightarrow ff})^{-1}.
\end{eqnarray}   
The singlet scalar must decay before the onset of BBN; otherwise, its late-time decay would disrupt the successful predictions of primordial nucleosynthesis.

%%%%%%%%%%%%%%%%%%%%%%%%%%%%%%%%%%%%%%%%%%%%%%%%%%%%%%%%%%%%%%%%%%%%%%%%%%
\subsection{Constraint from Higgs invisible decay}
\noindent
If the dark sector particles $X_D$ and $\phi$ are lighter than $M_h/2$, the SM Higgs boson can decay into pairs of these particles, contributing to its invisible decay width. Recent LHC measurements constrain the invisible branching ratio to $\mathcal{B}r(h\rightarrow \rm inv)\le0.145$ \cite{ATLAS:2022yvh}.
\begin{figure}[h]
\centering
\includegraphics[width=0.48\textwidth]{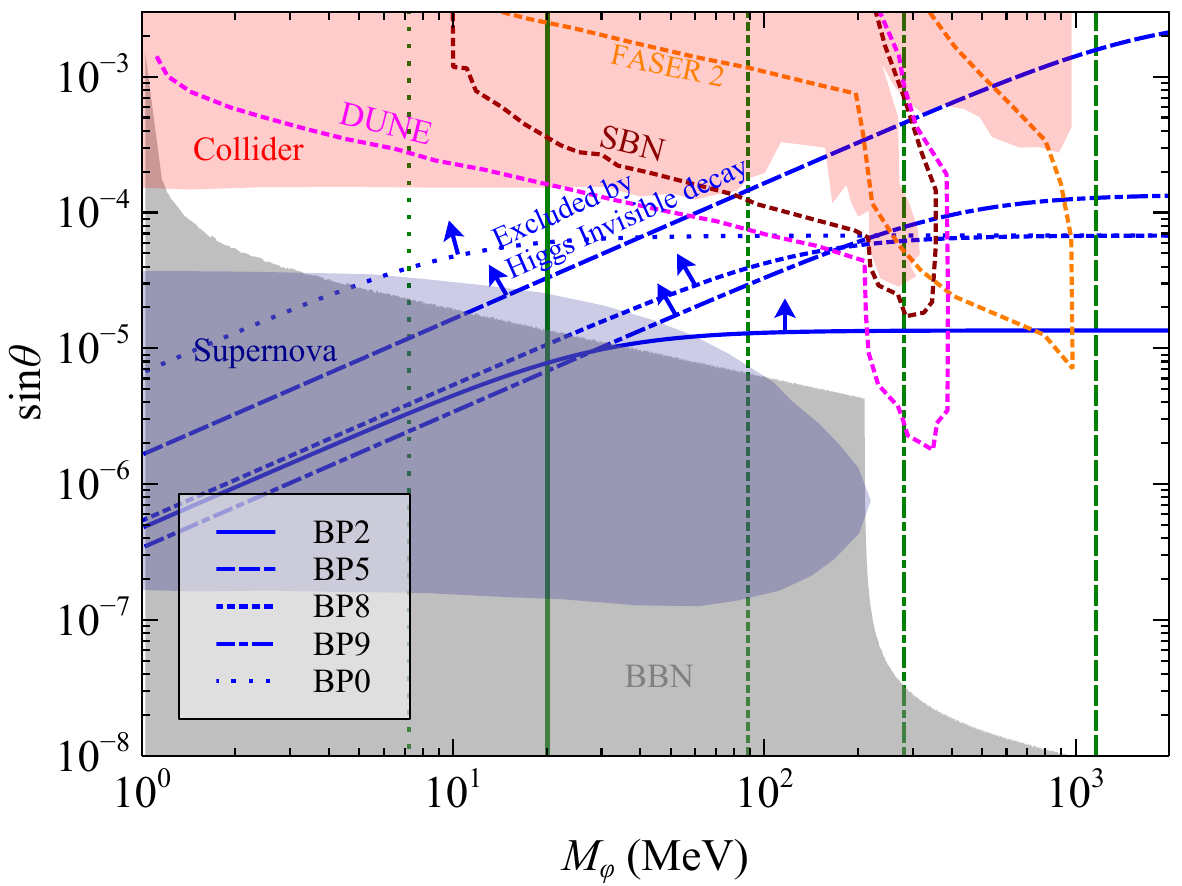}
\caption{Allowed parameter space from BBN and Higgs invisible decay constraints in the $\sin\theta$–$M_\phi$ plane. The gray shaded region is excluded because $\tau_\phi > \tau_{\rm BBN}$. The regions above the blue lines are excluded by Higgs invisible decay bounds. The constraint from supernova \cite{Balaji:2022noj} is shown with blue shaded region. The collider limits are shown with red shaded region \cite{Dev:2017dui,Egana-Ugrinovic:2019wzj,Dev:2019hho}. The future sensitivities from DUNE \cite{Berryman:2019dme}, SBN \cite{Batell:2019nwo}, and FASER 2 \cite{Anchordoqui:2021ghd} are shown with different colored dotted lines.}
\label{fig:bbn}
\end{figure}
The partial widths for Higgs decays into $\phi$ and $X_D$ pairs are
\begin{eqnarray}  \Gamma_{h\rightarrow\phi\phi}=\frac{\lambda_{H\Phi}^2v_h^2}{8\pi M_h}\sqrt{1-\frac{4M_\phi^2}{M_h^2}},
\end{eqnarray}
\begin{eqnarray}
    \Gamma_{h\rightarrow X_D X_D}&=&\frac{\textsl{g}_D^2 M_h^3\sin^2\theta}{8\pi M^2_{X_D}}\sqrt{1-\frac{4M_{X_D}^2}{M_h^2}}\nonumber\\&&\times\left( 1- \frac{4M_{X_D}^2}{M_h^2}+\frac{12M_{X_D}^4}{M_h^4}\right),
\end{eqnarray}
where $\lambda_{H\Phi}$ is the Higgs–portal coupling defined in Eq.~\eqref{eq:lamhphi}.

In Fig.~\ref{fig:bbn}, we show the combined BBN and Higgs invisible decay constraints in the $\sin\theta$–$M_\phi$ plane. The gray shaded region corresponds to $\tau_\phi > \tau_{\rm BBN} = 1~\text{s}$ and is therefore excluded by BBN. The constraints from supernova \cite{Balaji:2022noj} and colliders \cite{Dev:2017dui,Egana-Ugrinovic:2019wzj,Dev:2019hho} are shown with blue and red shaded regions, respectively. The future sensitivities of DUNE \cite{Berryman:2019dme}, SBN \cite{Batell:2019nwo}, and FASER 2 \cite{Anchordoqui:2021ghd} are shown by magenta, dark red, and orange dotted lines. The blue curves represent the Higgs invisible decay bounds for the five benchmark points (BP0, BP2, BP5, BP8, and BP9) in Tables~\ref{tab:tab1fa} and \ref{tab:tab1ff}. For each benchmark, we fix $\textsl{g}_D$ and $M_{X_D}$ to their benchmark values and vary $\{M_\phi,\sin\theta\}$, such that the region below each blue line satisfies $\mathcal{B}r(h\rightarrow {\rm inv}) \le 0.145$, while the region above is excluded. The values of $M_\phi$ corresponding to \textit{BP0}, \textit{BP2}, \textit{BP5}, \textit{BP8}, and \textit{BP9} are indicated by green spaced-dotted,  solid, dashed, dotted, and dashed-dotted vertical lines, respectively. Evaluating the viability of our benchmark points against this comprehensive set of limits, we find that \textit{BP1}, \textit{BP3}, \textit{BP4}, and \textit{BP7} are firmly excluded by the combined BBN and Higgs invisible decay constraints. Similarly, \textit{BP2} is ruled out by both the Higgs invisible decay bounds and supernova constraints. Furthermore, although \textit{BP9} safely evades the BBN and Higgs limits, it is ultimately excluded by existing collider bounds. Consequently, only \textit{BP0}, \textit{BP5}, \textit{BP6}, \textit{BP8}, and \textit{BP10} successfully survive all current experimental and cosmological constraints, establishing them as highly viable targets for future planned facilities.

%%%%%%%%%%%%%%%%%%%%%%%%%%%%%%%%%%%%%%%%%%%%%%%%%%%%%%%%%%%%%%%%%%%%%%%%%%
\section{Gravitational wave from first-order phase transition}\label{sec:gwfopt}
\noindent
Having established the viability of our benchmark points against various constraints, we now turn to the gravitational wave signals generated by the dark sector phase transition. The stochastic gravitational wave background arising from a strong first-order phase transition is shaped by contributions from bubble collisions, sound waves in the plasma, and magnetohydrodynamic turbulence, as detailed in Appendix~\ref{app:gw}. Here, we compute the GW spectrum for the five benchmark points (BP0, BP5, BP6, BP8, and BP10), that are allowed from all constraints, listed in Tables~\ref{tab:tab1fa} and \ref{tab:tab1ff} and compare it against the projected sensitivities of existing and planned detectors.

\begin{figure}[h]
\centering
\includegraphics[width=0.48\textwidth]{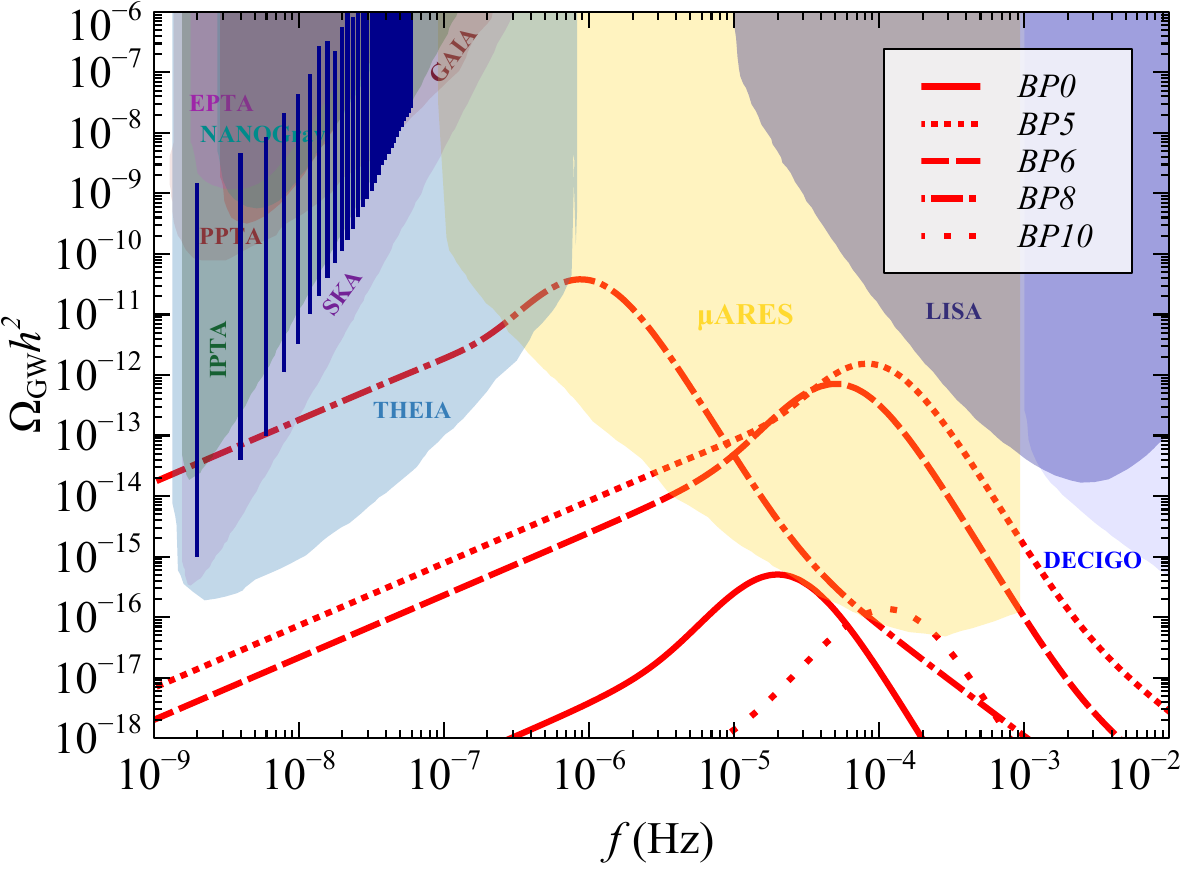}
\caption{Gravitational wave spectrum for the five benchmark points listed in Tables~\ref{tab:tab1fa} and \ref{tab:tab1ff}. The colored shaded regions indicate the projected sensitivities of different gravitational wave detectors.}\label{fig:gw}
\end{figure}

In Fig.~\ref{fig:gw}, we present the predicted stochastic gravitational wave spectra expressed as the GW energy density amplitude $\Omega_{\rm GW} h^2$ versus frequency $f$ for the selected benchmark points. For context, we overlay the current and projected sensitivities of a wide array of GW observatories, including NANOGrav \cite{NANOGrav:2023gor,NANOGrav:2023hvm}, EPTA \cite{EPTA:2023fyk}, PPTA \cite{Reardon:2023gzh}, IPTA \cite{Hobbs:2009yy}, SKA \cite{Weltman:2018zrl}, LISA \cite{LISA:2017pwj}, THEIA \cite{Garcia-Bellido:2021zgu}, GAIA \cite{Garcia-Bellido:2021zgu}, DECIGO \cite{Seto:2001qf}, and $\mu$ARES \cite{Sesana:2019vho}. The vertical blue lines denote the frequency and amplitude range currently favored by the NANOGrav 15-year dataset. The properties of these GW signals depend sensitively on the underlying dark sector dynamics. For instance, the MeV-scale DM mass of \textit{BP8} dictates a nucleation temperature $T_n$ that is correspondingly in the MeV range (cf.~Table~\ref{tab:tab1ff}). This phase transition scale naturally positions the resulting GW signal within the nHz to $\mu$Hz frequency bands, placing it within the observational reach of pulsar timing arrays (PTAs), SKA, $\mu$ARES, and astrometric observatories like THEIA. Notably, the low-frequency tail of this spectrum marginally intersects the parameter space indicated by the NANOGrav observations~\cite{NANOGrav:2023gor}, demonstrating that such MeV-scale transitions can begin to probe regions of current observational interest. Conversely, while \textit{BP0} also features an MeV-scale DM mass and $T_n$ where the relic density is dominantly decided by the $X_D \phi$ channel, its phase transition is characterized by a substantially larger inverse duration parameter ($\beta_*/\mathcal{H}_*$). This rapid transition severely suppresses the overall GW amplitude and shifts the peak frequency to $\sim 2\times10^{-5}$~Hz, placing it only at the very edge of the projected $\mu$ARES sensitivity. Finally, moving to the heavier mass regime, the GeV-scale DM masses characterizing \textit{BP5}, \textit{BP6}, and \textit{BP10} necessitate significantly higher phase transition scales ($T_n \sim \mathcal{O}(\rm GeV)$). Consequently, their GW peak frequencies are shifted substantially upward to $\sim 10^{-4}$~Hz, bringing them into the optimal sensitivity windows of $\mu$ARES and LISA.

A distinctive feature of this scenario is the strong correlation between the DM mass $M_\chi$ and the nucleation temperature $T_n$ of the phase transition. This connection arises naturally from the fundamental kinematic requirements of our framework: to maintain the dominant tree-level $X_D\phi$ annihilation channel in the strictly forbidden regime ($M_{X_D} + M_\phi > 2M_\chi$), the dark sector masses $M_{X_D}$ and $M_\phi$ must necessarily scale upward with the DM mass $M_\chi$. Since these masses are generated via spontaneous symmetry breaking ($M_{X_D} = \textsl{g}_D v_\phi$ and $M_\phi = \sqrt{2\lambda_\Phi} v_\phi$), achieving heavier final states without pushing the gauge and quartic couplings into non-perturbative regimes requires a larger scalar VEV, $v_\phi$, for heavier DM. Elevating the VEV inherently elevates the entire thermal scale of the dark sector. Consequently, both the critical temperature $T_c$ and the nucleation temperature $T_n$ generally increase with $M_\chi$, as is evident across the benchmark points spanning the MeV to GeV regimes in Table~\ref{tab:tab1fa} and Table~\ref{tab:tab1ff}. This robust model prediction that heavier secluded DM is intrinsically accompanied by higher-scale phase transitions and correspondingly higher-frequency GW signals provides a powerful consistency check for the scenario and establishes a clear target for near-future multi-frequency GW observations.

The detection prospects of the gravitational wave signal at a given observatory can be effectively quantified by the signal-to-noise ratio (SNR), defined as \cite{Schmitz:2020syl}
\begin{equation}
\label{eq:snr}
    \mathcal{S}=\sqrt{\tau_{\rm obs}\int_{f_{\rm min}}^{f_{\rm max}}df\left[\frac{\Omega_{\rm GW}h^2}{\Omega_{\rm expt}h^2}\right]^2}
\end{equation}
where $\tau_{\rm obs}$ is the total observation time and $\Omega_{\rm expt}h^2$ represents the noise sensitivity spectrum of the specific detector. 

In Fig.~\ref{fig:relicpar_scan1}, we overlay SNR contours corresponding to a detection threshold of $\mathcal{S}=10$ (assuming $\tau_{\rm obs}=1$ year) for BBO (cyan dashed), DECIGO (cyan solid), LISA (red solid), and $\mu$ARES (yellow dotted) across the $M_\chi$ versus $\textsl{g}_D$ parameter space. For a given $M_\chi$, the region to the right of each respective contour yields an SNR strictly greater than 10. The physical origin of this behavior is straightforward: increasing the dark gauge coupling $\textsl{g}_D$ strengthens the first-order phase transition, resulting in a larger latent heat fraction $\alpha_*$. This stronger FOPT subsequently drives a larger gravitational wave amplitude $\Omega_{\rm GW}h^2$, elevating the resulting SNR according to Eq.~\ref{eq:snr}. However, we note that for DM mass, $M_\chi\gtrsim 23$ GeV, we observe the same SNR, $\mathcal{S}=10$, for two different $\textsl{g}_D$ values for any fixed DM mass. This feature is a consequence of imposing the correct relic density criteria as explained in Sec~\ref{sec:fdmfopt}.

\begin{figure}[]
\centering
\includegraphics[width=0.48\textwidth]{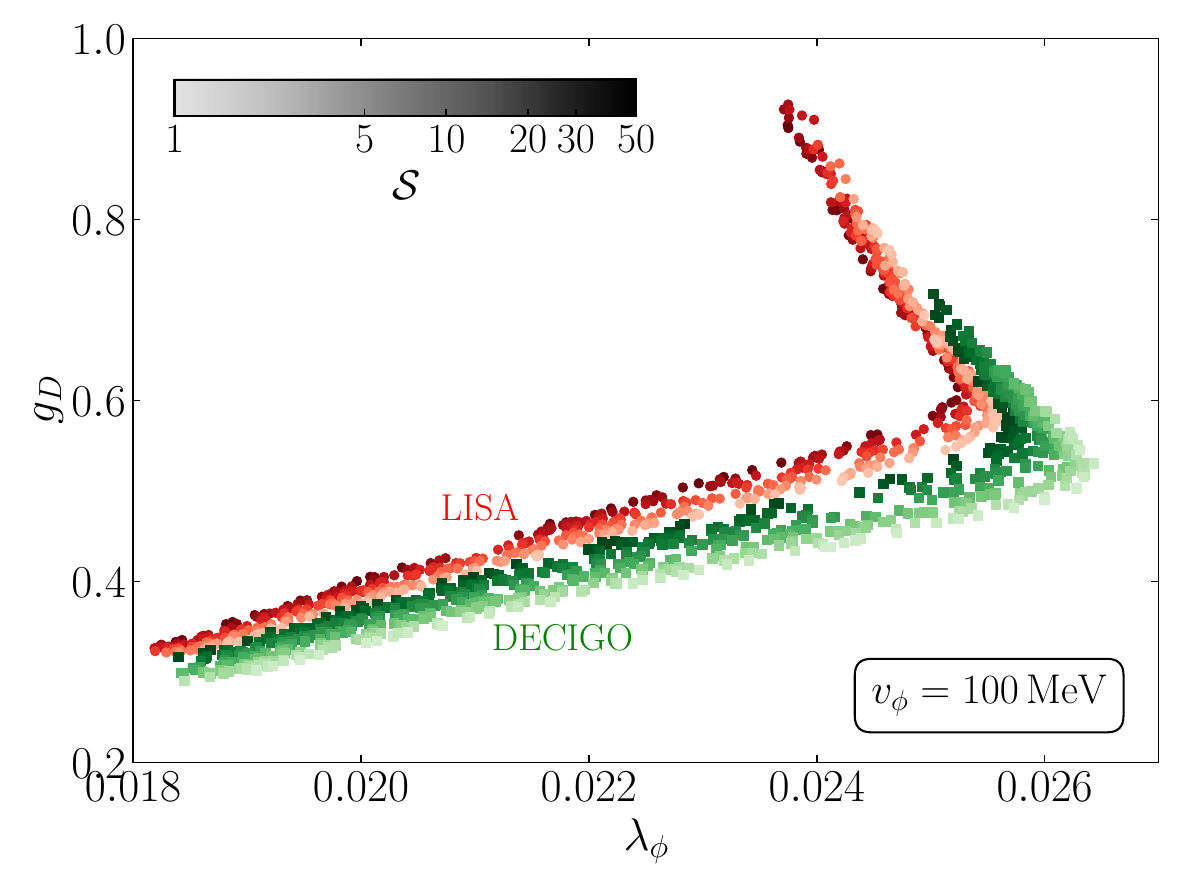}
\caption{FOPT parameter space consistent with DM relic abundance in the plane of $\textsl{g}_D$ vs $\lambda_\phi$. Red color represents LISA SNR, and the green color represents DECIGO SNR. The color grading denotes SNR values ranging from 1 to 50.}
\label{fig:snr100MeV}
\end{figure}

To further show the correlation of the phase transition dynamics and the relic density parameter space, we isolate the FOPT parameter space in Fig.~\ref{fig:snr100MeV}, mapping the exact SNR values across the $\textsl{g}_D$ versus $\lambda_\phi$ plane for the points previously validated in Fig.~\ref{fig:relicpar_scan1}. 
Here, we highlight parameter configurations yielding an SNR between 1 and 50, using a red color gradient for LISA and a green color gradient for DECIGO, with darker shades indicating higher SNR. As expected, a larger $\textsl{g}_D$ enhances the GW amplitude and pushes the SNR toward higher values. Notably, the LISA SNR band is offset and lies at higher $\textsl{g}_D$ values compared to the DECIGO band. Because LISA's projected strain sensitivity is lower than DECIGO's in this specific frequency range ($\Omega^{\rm LISA}_{\rm expt}h^2 > \Omega^{\rm DECIGO}_{\rm expt}h^2$), achieving the same signal-to-noise ratio requires a correspondingly stronger gravitational wave signal ($\Omega^{\rm LISA}_{\rm GW}h^2 > \Omega^{\rm DECIGO}_{\rm GW}h^2$). This strictly necessitates a stronger phase transition, thereby demanding a systematically larger gauge coupling $\textsl{g}_D^{\rm LISA} > \textsl{g}_D^{\rm DECIGO}$. Ultimately, the presence of these distinct, highly sensitive SNR bands highlights the robust testability of our framework. We note that the SNR bands for BBO and $\mu$ARES are very similar to those of DECIGO and LISA, respectively, as can be seen from Fig. \ref{fig:relicpar_scan1}. Therefore, we refrain from showing BBO and $\mu$ARES SNR bands in Fig. \ref{fig:snr100MeV}. Future detection of this stochastic gravitational wave background by observatories such as LISA or DECIGO would not merely signify a cosmological phase transition but could also pinpoint the underlying microscopic gauge and scalar dynamics of the dark sector. When combined with ongoing direct detection and collider searches, these gravitational wave signatures offer a powerful, complementary avenue for decisively probing the parameter space of this forbidden DM scenario.

%%%%%%%%%%%%%%%%%%%%%%%%%%%%%%%%%%%%%%%%%%%%%%%%%%%%%%%%%%%%%%%%%%%%%%%%%%
\section{Conclusions}\label{sec:concl}
\noindent
We have presented a minimal $U(1)_D$ gauge extension of the Standard Model featuring a Dirac fermion dark matter candidate $\chi$ and a complex singlet scalar $\Phi$. The spontaneous breaking of the $U(1)_D$ symmetry is absolutely indispensable for the viability of the scenario: in its absence, the dark gauge boson $X_D$ and the scalar $\phi$ would remain massless, leading to unsuppressed annihilations that would severely deplete the DM relic abundance and catastrophically violate late-time cosmological bounds. The symmetry breaking elegantly resolves this by serving a dual purpose: it generates masses for both $X_D$ and $\phi$ to reshape the annihilation kinematics, and it simultaneously triggers a strong first-order phase transition in the dark sector.

A defining feature of this model is the specific $U(1)_D$ charge assignment, which strictly forbids the standard $\bar{\chi}\chi \rightarrow \phi\phi$ annihilation at tree level, restricting it to a velocity-dependent $p$-wave loop process. To achieve the correct relic abundance, we restrict our parameter space such that the dominant tree-level $s$-wave annihilation, $\bar{\chi}\chi \rightarrow X_D \phi$, is strictly kinematically forbidden at zero temperature. Consequently, we find that the DM relic density is determined by two distinct regimes: (i) when the $\bar{\chi}\chi \rightarrow \phi\phi$ channel is kinematically allowed (creating a rich interplay between a forbidden tree-level process and an allowed loop-level $p$-wave process) and 
(ii) when both the $\bar{\chi}\chi \rightarrow X_D \phi$ and $\bar{\chi}\chi \rightarrow \phi\phi$ channels are kinematically forbidden.

Crucially, our scenario comfortably evades stringent constraints from CMB observations and indirect detection experiments through two distinct protective mechanisms: the energetic threshold fundamentally shuts off the forbidden $\bar{\chi}\chi \rightarrow X_D\phi$ channel at late times, while the explicit $p$-wave nature of the loop-induced $\bar{\chi}\chi \rightarrow \phi\phi$ channel renders its late-time cross-section vanishingly small. The benchmark points that satisfy the observed relic density also safely navigate existing direct detection limits, Big Bang Nucleosynthesis bounds, and Higgs invisible decay constraints, while remaining highly testable by near-future low-threshold experiments.

Furthermore, the FOPT driving this dark sector mechanism produces a stochastic gravitational wave background spanning the nHz to mHz frequency range, placing the signal within reach of existing pulsar timing arrays (NANOGrav, EPTA, PPTA, IPTA) and future detectors, including SKA, THEIA, $\mu$ARES, LISA, and DECIGO. We uncover a robust correlation between the DM mass $M_\chi$ and the nucleation temperature $T_n$: maintaining the necessary kinematic threshold ($M_{X_D} + M_\phi > 2M_\chi$) for heavier DM requires a correspondingly larger scalar VEV, thereby elevating the phase transition temperature scale. This dynamically shifts the GW peak frequency upward for heavier DM candidates. Taken together, this framework demonstrates that a rigorous connection between DM freeze-out and early Universe phase transitions can simultaneously address the relic density puzzle and yield rich, mutually reinforcing experimental signatures across multiple observational frontiers.

%%%%%%%%%%%%%%%%%%%%%%%%%%%%%%%%%%%%%%%%%%%%%%%%%%%%%%%%%%%%%%%%%%%%%%%%%%
\noindent
\acknowledgments{P.K.P. would like to acknowledge the Ministry of Education, Government of India, for providing financial support for his research via the Prime Minister’s Research Fellowship (PMRF) scheme. Parts of this work were developed through discussions among the authors at the \emph{Phoenix 2025} conference, and we thank the organizers for providing a stimulating environment. S.M. acknowledges support from the IIT Goa Startup Grant [2025/SG/SM/057].}

%%%%%%%%%%%%%%%%%%%%%%%%%%%%%%%%%%%%%%%%%%%%%%%%%%%%%%%%%%%%%%%%%%%%%%%%%%
\appendix
%%%%%%%%%%%%%%%%%%%%%%%%%%%%%%%%%%%%%%%%%%%%%%%%%%%%%%%%%%%%%%%%%%%%%%%%%%
\section{SM Higgs and $\Phi$ mixing}
\noindent
The scalar potential of the model containing the singlet scalar $\Phi$ and the SM Higgs doublet $H$ is given by
    \begin{eqnarray}
        V(H, \Phi)&=&-\mu^2_\Phi \Phi^2 -\mu_H^2 (H^\dagger H)  + \lambda_H(H^\dagger H)^2 + \lambda_{\Phi}\Phi^4\nonumber\\&&+\lambda_{H\Phi}(H^\dagger H)\Phi^2.
        \label{Eq:scalarpot}
    \end{eqnarray}
Parameterizing the scalars $H$ and $\Phi$ as
    \begin{equation}
        H=\frac{1}{\sqrt{2}}\begin{pmatrix}
            0\\
            v_h+h
        \end{pmatrix}\,,~~ \Phi=\frac{\phi+v_\phi}{\sqrt{2}}\label{eq:physicalscalar}
    \end{equation}
    where the VEVs are estimated to be $$v_h=\sqrt{ \frac{2 \mu_H^2-\lambda_{H\Phi}v_\phi^2}{2\lambda_H}}, ~v_\phi=\sqrt{ \frac{2 \mu_\Phi^2-\lambda_{H\Phi}v_h^2}{2\lambda_\Phi}}.$$
    The mass-squared matrix spanning the $h$ and $\phi$ is given by
    \begin{equation}
        M^2_{h\phi}=\begin{pmatrix}
            2\lambda_H v_h^2 & \lambda_{H\Phi}v_hv_\phi\\
            \lambda_{H\Phi}v_hv_\phi &  2\lambda_{\Phi}v_\phi^2
        \end{pmatrix}.
    \end{equation}
    Diagonalizing the above mass matrix, we get the mass eigenstates $h_1$ with mass  $M_{h_1}=125$ GeV and $h_2$ with mass $M_{h_2}$. The $h-\phi$ mixing angle is then given by  
    \begin{equation}
        \tan 2\theta \sim \frac{\lambda_{H\Phi}v_hv_\phi}{\lambda_H v_h^2-\lambda_{\Phi}v_\phi^2}.\,
    \end{equation}

The scalar couplings can be expressed in terms of the scalar masses and mixing angle as
\begin{eqnarray}
\lambda_H=\frac{\cos\theta^2M_{h_1}^2+\sin\theta^2M_{h_2}^2}{2v_h^2},
\end{eqnarray}
\begin{eqnarray}
\lambda_\Phi=\frac{\cos\theta^2M_{h_2}^2+\sin\theta^2M_{h_1}^2}{2v_\phi^2},
\end{eqnarray}
\begin{eqnarray}
\lambda_{H\Phi}=\cos\theta\sin\theta \frac{M_{h_1}^2-M_{h_2}^2}{v_h v_\phi}.\label{eq:lamhphi}
\end{eqnarray}

%%%%%%%%%%%%%%%%%%%%%%%%%%%%%%%%%%%%%%%%%%%%%%%%%%%%%%%%%%%%%%%%%%%%%%%%%%%%%%%%%%%%%%%%%%%%%%%%%%%%%%%%%%%%%%%
\section{DM annihilation and scattering cross-sections}
\noindent
We used \texttt{CalcHEP} \cite{Belyaev:2012qa} for computing the cross-sections after implementing the model in \texttt{Lanhep} \cite{Semenov:2014rea}. The thermally averaged cross-section for any process $12\rightarrow34$ is given as
\begin{eqnarray}
\langle\sigma v\rangle_{12\rightarrow34}&=&\frac{T}{8m_1^2m_2^2K_2(\frac{m_1}{T})K_2(\frac{m_2}{T})}\times\nonumber\\&&\int_{{\rm max}[(m_1+m_2)^2,(m_3+m_4)^2]}^\infty ds~\sigma(s)\sqrt{s}K_1\bigg(\frac{\sqrt{s}}{T}\bigg)\nonumber\\&&\times\big(s-{\rm max}[(m_1+m_2)^2,(m_3+m_4)^2]\big).
\end{eqnarray}

%%%%%%%%%%%%%%%%%%%%%%%%%%%%%%%%%%%%%%%%%%%%%%%%%%%%%%%%%%%%%%%%%%%%%%%%%%%%%%%%%%%%%%%%%%%%%%%%%%%%%%%%%%%%%%%
\section{Details of first-order phase transition}\label{app:fopt}
\noindent
The finite temperature effective potential is given as
\begin{eqnarray}
V_{\rm eff}(\phi,T)&=&V_{\rm tree}(\phi)+V_{\rm CW}(\phi)+V_{\rm ct}(\phi)+V_{T}(\phi,T)\nonumber\\&&+V_{\rm daisy}(\phi,T).
\end{eqnarray}
The tree-level part of the potential is given as
\begin{eqnarray}
 V_{\rm tree}=-\frac{1}{2}\lambda_{\Phi}v_\phi^2\phi^2+\frac{1}{4}\lambda_{\Phi}\phi^4.
\end{eqnarray}
The one-loop Coleman-Weinberg zero temperature contribution in $\rm \overline{MS}$ renormalization scheme is given as
\begin{eqnarray}
V_{\rm CW}&=&\frac{1}{64\pi^2}m^4_\phi(\phi)\bigg(\log\bigg(\frac{m_\phi^2(\phi)}{\mu_R^2}\bigg)-\frac{3}{2}\bigg)\nonumber\\&&+\frac{3}{64\pi^2}m^4_{X_D}(\phi)\bigg(\log\bigg(\frac{m_{X_D}^2(\phi)}{\mu_R^2}\bigg)-\frac{5}{6}\bigg),
\end{eqnarray}
where the field-dependent masses are given as
\begin{eqnarray}
m_{\phi}(\phi)=\sqrt{-\lambda_{\Phi}v^2+3\lambda_{\Phi}\phi^{2}}~;
M_{Z_{D}}(\phi)=\textsl{g}_D\phi.
\end{eqnarray}
The renormalization scale is fixed at $\mu_R\equiv v_\phi$. The counter term is
\begin{eqnarray}
V_{\rm ct}=-\frac{\delta\mu_\Phi^2}{2}\phi^2+\frac{\delta\lambda_{\Phi}}{4}\phi^4,
\end{eqnarray}
which is derived by solving
\begin{eqnarray}
\frac{\partial(V_{\rm CW}+V_{\rm ct})}{\partial\phi}\bigg|_{\phi=v_\phi}=0,\frac{\partial^2(V_{\rm CW}+V_{\rm ct})}{\partial\phi^2}\bigg|_{\phi=v_\phi}=0.
\end{eqnarray}
The finite temperature contribution is given as
\begin{eqnarray}
V_{T}&=&\frac{T^4}{2\pi^2}J_{B}\bigg(\frac{m_\phi(\phi)}{T}\bigg)+\frac{T^4}{2\pi^2}J_{B}\bigg(\frac{m_\eta(\phi)}{T}\bigg)\nonumber\\&&+3\frac{T^4}{2\pi^2}J_{B}\bigg(\frac{m_{X_D}(\phi)}{T}\bigg),
\end{eqnarray}
where, $m_{\eta}(\phi)=\sqrt{-\lambda_{\phi}v^2+\lambda_{\phi}\phi^{2}}$ and 
\begin{eqnarray}
J_{B}(y)=\int_0^{\infty}x^2\log\big(1- e^{-\sqrt{x^2+y^2}}\big).
\end{eqnarray}
The daisy contribution is 
\begin{eqnarray}
V_{\rm daisy}=-\frac{\textsl{g}_D^3T}{12\pi}\bigg( \Big(\phi^2+T^2\Big)^{3/2}-\phi^3  \bigg).
\end{eqnarray}
The bubble nucleation rate per unit volume at a finite temperature is given by
\begin{eqnarray}
\Gamma(T)\simeq \bigg(\frac{S_3(T)}{2\pi T}\bigg)^{3/2}T^4 e^{-\frac{S_3(T)}{T}},
\end{eqnarray}
where $S_3$ is the three-dimensional Euclidean action given as 
\begin{eqnarray}
S_3(T)=\int d^3r\bigg[\frac{1}{2}\big(\frac{d\phi}{dr}\big)^2+V_{\rm eff}(\phi,T)\bigg],
\end{eqnarray}
which is calculated by solving
\begin{eqnarray}
\frac{d^2\phi}{dr^2}+\frac{2}{r}\frac{d\phi}{dr}-\frac{dV_{\rm eff}(\phi,T)}{d\phi}=0,
\end{eqnarray}
with the following boundary conditions
\begin{eqnarray}
\frac{d\phi}{dr}\bigg|_{r=0}=0, ~\phi(\infty)=\phi_{\rm false}.
\end{eqnarray}
The nucleation process begins when the bubble nucleation rate becomes comparable to the Hubble expansion rate, which is described by
\begin{eqnarray}
\Gamma(T_n)\approx \mathcal{H}(T_n)^4,
\end{eqnarray}
where $T_n$ is the nucleation temperature.

%%%%%%%%%%%%%%%%%%%%%%%%%%%%%%%%%%%%%%%%%%%%%%%%%%%%%%%%%%%%%%%%%%%%%%%%%%
\section{Gravitational waves}\label{app:gw}
\textbf{(i) Bubble collision:}

The peak frequency and the amplitude of the GWs generated by the bubble collision are given by
\begin{eqnarray}
f^{{\rm col}}_{\rm peak}&=&1.65\times10^{-5}({\rm Hz})\bigg(\frac{g_{*}(T_n)}{100}\bigg)^{1/6}\bigg(\frac{T_n}{0.1 {~\rm TeV}}\bigg)\nonumber\\&&\frac{0.64}{2\pi}\bigg(\frac{\beta(T_n)}{\mathcal{H}(T_n)}\bigg),
\end{eqnarray}

\begin{eqnarray}
\Omega_{{\rm col}}h^2&=&1.65\times10^{-5}\bigg(\frac{100}{g_{*}(T_n)}\bigg)^{1/3}\bigg(\frac{\mathcal{H}(T_n)}{\beta(T_n)}\bigg)\nonumber\\&&\bigg(\frac{\kappa_\phi(T_n)\alpha(T_n)}{1+\alpha(T_n)}\bigg)^2 \frac{A(a+b)^c}{\bigg(b\big(\frac{f}{f^{{\rm col}}_{\rm peak}}\big)^{-a/c}+a\big(\frac{f}{f^{{\rm col}}_{\rm peak}}\big)^{b/c}\bigg)^c},\nonumber\\
\end{eqnarray}
with the efficiency factor 
\begin{eqnarray}
\kappa_\phi=\frac{1}{1+0.715\alpha(T_n)}\bigg(0.715\alpha(T_n)+\frac{4}{27}\sqrt{\frac{3\alpha(T_n)}{2}}\bigg),\nonumber\\
\end{eqnarray}
and $a=1.03,b=1.84,c=1.45,A=5.93\times10^{-2}$ \cite{Lewicki:2022pdb, Athron:2023xlk, Caprini:2024hue}.

\noindent
\textbf{(ii) Sound waves:}

The peak frequency and the amplitude of the GWs generated by the sound waves are expressed by \cite{Athron:2023xlk} 
\begin{eqnarray}
f^{{\rm sw}}_{\rm peak}&=&8.9\times10^{-6}({\rm Hz})\bigg(\frac{g_{*}(T_n)}{100}\bigg)^{1/6}\bigg(\frac{1}{v_w}\bigg)\nonumber\\&&\bigg(\frac{T_n}{0.1 {~\rm TeV}}\bigg)\bigg(\frac{z_p}{10}\bigg)\bigg(\frac{\beta(T_n)}{\mathcal{H}(T_n)}\bigg),
\end{eqnarray}
where the wall velocity $v_w\sim1$, and $z_p\sim10$.

\begin{eqnarray}
\Omega_{{\rm sw}}h^2&=&2.59\times10^{-6}\bigg(\frac{100}{g_{*}(T_n)}\bigg)^{1/3}\bigg(\frac{\mathcal{H}(T_n)}{\beta(T_n)}\bigg)v_w\Upsilon(T_n) \nonumber\\&&\bigg(\frac{\kappa_{\rm sw}(T_n)\alpha(T_n)}{1+\alpha(T_n)}\bigg)^2 \frac{7^{3.5}\big(\frac{f}{f^{{\rm sw}}_{\rm peak}(T_n)}\big)^3}{\bigg(4+3\big(\frac{f}{f^{{\rm col}}_{\rm peak}(T_n)}\big)^{2}\bigg)^{3.5}}.
\end{eqnarray}
The efficiency factor is \cite{Espinosa:2010hh}
\begin{eqnarray}
\kappa_{{\rm sw}}=\frac{\alpha(T_n)}{0.73+0.083\sqrt{\alpha(T_n)}+\alpha(T_n)}
\end{eqnarray}
The suppression factor is given as
\begin{eqnarray}
\Upsilon=1-\frac{1}{\sqrt{1+2\tau_{{\rm sw}}(T_n)\mathcal{H}(T_n)}},
\end{eqnarray}
where the lifetime of the sound wave is given as \cite{Guo:2020grp}
\begin{eqnarray}
\tau_{{\rm sw}}=\frac{R_{*}(T_n)}{U_f(T_n)},
\end{eqnarray}
where mean bubble separation is given by
\begin{eqnarray}
R_{*}=\frac{(8\pi)^{1/3}v_w}{\beta(T_n)},
\end{eqnarray}
and the rms fluid velocity is given as
\begin{eqnarray}
U_f=\sqrt{\frac{3\kappa_{{\rm sw}}(T_n)\alpha(T_n)}{4(1+\alpha(T_n))}}.
\end{eqnarray}

\noindent
\textbf{(iii) Turbulence:}

The peak frequency and the amplitude of the GWs generated by the turbulence are expressed by \cite{Caprini:2015zlo, Athron:2023xlk, Caprini:2024hue}
\begin{eqnarray}
f^{{\rm turb}}_{\rm peak}&=&2.7\times10^{-5}({\rm Hz})\bigg(\frac{g_{*}(T_n)}{100}\bigg)^{1/6}\bigg(\frac{1}{v_w}\bigg)\nonumber\\&&\bigg(\frac{T_n}{0.1 {~\rm TeV}}\bigg)\bigg(\frac{\beta(T_n)}{\mathcal{H}(T_n)}\bigg),
\end{eqnarray}

\begin{eqnarray}
\Omega_{{\rm turb}}h^2&=&3.35\times10^{-4}\bigg(\frac{100}{g_{*}(T_n)}\bigg)^{1/3}\bigg(\frac{\mathcal{H}(T_n)}{\beta(T_n)}\bigg)\times\nonumber\\&&\bigg(\frac{\kappa_{\rm turb}(T_n)\alpha(T_n)}{1+\alpha(T_n)}\bigg)^2 v_w \times\nonumber\\&&\frac{\big(\frac{f}{f^{{\rm sw}}_{\rm peak}(T_n)}\big)^3}{\bigg(1+\big(\frac{f}{f^{{\rm col}}_{\rm peak}(T_n)}\big)\bigg)^{3.6}\big(1+\frac{8\pi f}{h_*(T_n)}\big)},
\end{eqnarray}
where the efficiency factor is $\kappa_{{\rm turb}}=0.1\kappa_{{\rm sw}}$ \cite{Caprini:2015zlo}, and the inverse of the Hubble time at the epoch of GW production, red-shifted to today, is expressed as:
\begin{eqnarray}
h_*=1.65\times10^{-5}({\rm Hz})\bigg(\frac{g_{*}(T_n)}{100}\bigg)^{1/6}\bigg(\frac{T_n}{0.1 {~\rm TeV}}\bigg)
\end{eqnarray}

%%%%%%%%%%%%%%%%%%%%%%%%%%%%%%%%%%%%%%%%%%%%%%%%%%%%%%%%%%%%%%%%%%%%%%%%%%%%%%%%%%%%%%%%%%%%%%%%%%%%%%%%%%%%%%%


\end{document}